\documentclass[twocolumn,english,aps,prl,preprint,reprint,showpacs,longbibliography,showkeys]{revtex4-1}
\usepackage[T1]{fontenc}
\usepackage{lmodern}
\usepackage{color}
\usepackage{babel}
\usepackage{amsmath}
\usepackage{amssymb}
\usepackage{graphicx}
\renewcommand{\Re}{\mathrm{Re}}
\RequirePackage[colorlinks=true,linkcolor=blue]{hyperref}
\usepackage{dsfont}
\usepackage{cleveref}
\usepackage{natbib}
\usepackage{bm}
    
\makeatletter
\usepackage{amsfonts}
\usepackage{babel}
\usepackage{braket}
\usepackage{verbatim}
\usepackage{amstext}
\makeatother

\begin{document}

\title{Dissipative long-range entanglement generation between electronic spins}

\author{M. Benito,$^{1,2,*}$ M. J. A. Schuetz,$^{2,*}$ J. I. Cirac,$^{2}$ G. Platero,$^{1}$ and G. Giedke$^{2,3,4}$}

\address{$^{1}$Instituto de Ciencia de Materiales, CSIC, Sor Juana Ines de
la Cruz, 3, Cantoblanco, 28049 Madrid, Spain}

\address{$^{2}$Max-Planck-Institut f\"ur Quantenoptik, Hans-Kopfermann-Str.
1, 85748 Garching, Germany}

\address{$^{3}$Donostia International Physics Center, Paseo Manuel de
Lardizabal 4, 20018 Donostia-San Sebasti\'an, Spain}

\address{$^{4}$Ikerbasque, Basque Foundation for Science, Maria Diaz de Haro
3, 48013 Bilbao, Spain}

\thanks{These authors have contributed equally to this work.}

\begin{abstract}
We propose a scheme for deterministic generation and long-term stabilization
of entanglement between two electronic spin qubits confined in spatially
separated quantum dots. 
Our approach relies on an electronic quantum bus, consisting either of quantum Hall edge channels or surface acoustic waves, 
that can mediate long-range coupling between localized spins over distances of tens of micrometers. 
Since the entanglement is actively stabilized
by dissipative dynamics, our scheme is inherently robust against
noise and imperfections. 
\end{abstract}


\maketitle
%-----------------------------------------------------------------------------
\section{Introduction}
The physical realization of a large-scale quantum information processing (QIP) architecture constitutes a
fascinating problem at the interface between 
fundamental science and engineering \citep{HaAw08, NielsenandChuang}. 
Further advances towards this goal hinge upon two major challenges: 
(i) control over the undesired influences of the environment which tend to corrupt genuine quantum properties such as entanglement, and
(ii) long-range coupling between the logical qubits. 
The latter  not only relaxes some serious architectural challenges \citep{ScBl14} but also allows for applications in quantum communication, distributed quantum computing, and
some of the highest tolerances in 
error-correcting codes 
that are
based on long-distance entanglement links \citep{NielsenandChuang, knill05, nickerson13}.

In the solid state, electron spins confined in electrically defined semiconductor quantum dots (QDs) have emerged as a promising platform for QIP 
\citep{HKP+07,Kloeffel2013}:
Major building blocks such as initialization, single-shot readout, coherent control of single spins, and two-qubit gates between adjacent spins have been demonstrated successfully in proof-of-principle experiments. 
However, at present 
the integration of several qubits into a scalable architecture still remains a formidable challenge 
\citep{HKP+07, Braakman2013,Busl2013}.
A large amount of wiring and control electronics needs to be accommodated on a very small scale, since interactions between QDs are very short-range, enabling QIP setups with nearest-neighbor interactions only.
%, because (due to short-range interactions)
%a large amount of wiring and control electronics needs to be accommodated on a very small scale.
Therefore, a 
scalable design is likely to 
require \textit{long-range} couplings over distances of several micrometers \citep{ScBl14, ABD+13}.

In this work, we propose a scheme for
deterministic preparation of \textit{steady-state} entanglement between \textit{remote} qubits, defined by electron spins in spatially separated QDs. 
Our approach addresses the two challenges (i) and (ii) as described above within one unified framework: 
(i) By suitably engineering the continuous coupling of the system to its environment, our setup actively utilizes dissipation to create and stabilize quantum coherences, turning dissipation into the driving force behind the emergence of coherent quantum phenomena. 
This approach 
\citep{Plenio1999,Braun2002,Benatti2003, Muschik2011}
comes with potentially significant advantages over previous proposals \citep{Trifunovic2012,Trifunovic2013,Yang2015} which aim at a coherent coupling between remote spins, as dissipative methods are unaffected by timing and preparation errors and inherently robust against weak random perturbations, allowing us
to stabilize entanglement for arbitrary times \citep{Krauter2011, Sanchez2013, Schuetz2013, bohr15}.
(ii) Our scheme directly builds upon recent experimental developments towards the realization of a solid-state electronic quantum bus, where flying electrons take over the role of photons in more conventional atomic, molecular, and optical 
based approaches 
 in order to mediate long-range coupling between remote qubits. 
In particular, we consider
quantum Hall edge (QHE) channels 
\citep{Komiyama1992, Ji2003, Stace2004,  Roulleau2008a, Bocquillon2014, Thalineau2014, Yang2015} 
and surface acoustic waves (SAWs) 
\citep{wixforth89, Barnes2000, Stotz2005, Hermelin2011, McNeil2011, SKG+13, Bertrand} as exemplary candidate systems for the coherent transport of electron spins over long distances. 
Intuitively, the dissipative entanglement creation arises from a quantum interference effect in the \textit{common} coupling of the localized spins $\mathbf{S}_{i} (i=1,2)$ to an adjacent electronic quantum channel, in which flying electrons continuously pass by the two localized spins. 
With any which-way information absent, first-order spin-flip processes between the localized spins and the flying ancilla spins occurring in the course of electron transport can happen either in the first \textit{or} in the second node, which may lead to the formation of entanglement between the nodes, if two or more such processes with a unique common entangled steady-state dominate the dynamics \citep{TiVi12, Schuetz2013}. 
%\textcolor{blue}{Engineering of the electronic system parameters via external gate voltages then facilitates control over the desired steady-state properties.}

This work is structured as follows. 
In Sec.~\ref{sec:dissip} we introduce two generic \textit{dissipative} entanglement-generating dynamics, with a subsequent discussion on the robustness inherent to dissipative state preparation schemes. 
In Sec.~\ref{sec:model} we then propose and analyze two different physical setups, based on (i) QHE channels (see Sec.~\ref{subsec:QHE}) and (ii) SAW-induced moving quantum dots (see Sec.~\ref{subsec:SAW}), in order to approximately implement the paradigmatic schemes discussed in Sec.~\ref{sec:dissip}. 
In Sec.~\ref{sec:results} we turn to the central question of whether the steady-state entanglement found for the idealized dynamics can prevail in a realistic, noisy scenario.  
We discuss the dominant error sources, specify the experimental requirements, and provide a comprehensive comparison of the different setups. 
Finally, in Sec.~\ref{sec:conclusions} we draw conclusions and give an outlook on future directions of research. 

%-----------------------------------------------------------------------------
\section{Dissipative engineering}
\label{sec:dissip}
Let us first consider two different generic dissipative entanglement-generating dynamics for the system's density matrix (DM) $\rho$. 
A purely dissipative master equation (ME) with a unique entangled steady
state is given by \citep{Muschik2011}
\begin{equation}
\dot{\rho}=\alpha{\cal D}\left[\mu S_{1}^{+}+\nu S_{2}^{+}\right]\rho+\beta{\cal D}\left[\nu S_{1}^{-}+\mu S_{2}^{-}\right]\rho,\label{eq:goal1}
\end{equation}
where $S_{i}^{\pm},\ i=1,2$ denote the (spin) raising and lowering
operators for the two qubits and ${\cal D}\left[A\right]\rho=2A\rho A^{\dagger}-A^{\dagger}A\rho-\rho A^{\dagger}A$.
For all rates $\alpha,\beta{>}0$, the dissipative evolution given in Eq.~(\ref{eq:goal1}) drives the system into the
steady state
$\left|\Psi_{\text{ss}}\right>=\mu \left| {\uparrow}{\downarrow} \right>-\nu \left|{\downarrow}{\uparrow}\right>$,
which is unique and entangled for all $\mu,\nu{>}0,\ \mu{\neq}\nu$. 
While the entanglement is largest as $\mu\to\nu$, for equality the steady state is no longer unique (as is the case if one of the rates is zero). 
When there is more than one steady state, the long-time behavior depends on the initial state and may be strongly affected by small perturbations;
for example,
for $\beta=0$  (that is, for only one Lindblad term)
in Eq.~(\ref{eq:goal1}). Still,
a pure unique entangled steady state can be recovered by adding a suitable
Hamiltonian term \citep{Stannigel2012}, e.g.,
\begin{equation}
\dot{\rho}  =  -i\left[H,\rho\right]+\gamma{\cal D}\left[S_{1}^{+}+S_{2}^{+}\right]\rho,\label{eq:goal2}
\end{equation}
where $H =  2\Omega (S_{1}^{x}+S_{2}^{x})-i\Delta (S_{2}^{-}S_{1}^{+}-S_{2}^{+}S_{1}^{-})$, 
with $S_{i}^{x}=\left(S_{i}^{+}+S_{i}^{-}\right)/2$. 
Here, the corresponding (unnormalized) steady state reads
$\left|\Psi_{\text{ss}}\right>=\left|{\uparrow}{\uparrow}\right>+i\sqrt{2}\Omega/\Delta \left|{\cal S}\right>$,
where
$|{\cal S}\rangle=\left(\left|{\uparrow}{\downarrow}\right>-\left|{\downarrow}{\uparrow}\right>\right)/\sqrt{2}$ is the maximally entangled singlet state.

Our task in the following is
then to find or engineer an environment for two physical spins $\mathbf{S}_i$
that leads to the effective dynamics described by Eqs.~(\ref{eq:goal1}) or~(\ref{eq:goal2}).

\textit{Robustness.---}An important advantage of dissipative state preparation
schemes is their robustness, i.e., that the relevant qualitative and
quantitative features of the target state are preserved under
perturbations ${\cal L}_1$ of the dynamics. It is a feature of the
contractive dynamics generated by Lindblad-form Liouvillians that
the schemes are inherently unaffected by transient, timing, and
preparation errors; moreover, perturbations do not affect the
steady-state eigenvalue, which remains 0. Standard perturbation theory
(cf., e.g., \cite{Li2014,Benatti2011}) shows that the changes to the steady state (and to
the other eigenvalues) remain small (for a nondefective/nondegenerate
${\cal L}_0$) as long as $\alpha = \|{\cal L}_1\|$ (i.e., the strength
of the perturbation) is small compared to the smallest (in modulus)
nonzero eigenvalue of ${\cal L}_0$. This latter number is lower bounded
by the "dissipative" or "spectral" gap of ${\cal L}_0$, determined by the eigenvalue of the
Liouvillian with the largest real part different from zero, i.e.,
 $\epsilon=-\max \left\{\Re(\lambda_{i}) \right\} $,
where $\lambda_{i}$ are the nonzero eigenvalues of the Liouvillian.
\section{The model}
\label{sec:model}
In what follows, we show how our general idea can be applied to two different exemplary 
physical setups, with the ultimate goal of approximately
implementing the paradigmatic entanglement-generating dynamics given in Eqs.~(\ref{eq:goal1}) and (\ref{eq:goal2}), using a fermionic environment. 
First, we investigate QHE states as this setup facilitates direct analogies to existing quantum optical schemes with photons \citep{Bocquillon2014}. 
Thereafter, we explore a setup based on electrically induced SAWs where the stroboscopic control over the effective interaction times between stationary and mobile electron spins \citep{Hermelin2011, McNeil2011} results in larger amounts of entanglement. 
To treat each specific physical setup we employ two different input-output approaches tailored to the specific setups.

In all setups specified below, to controllably amplify the coupling between localized and flying electrons, we introduce auxiliary (ancilla) QDs that are tunnel-coupled to the QDs hosting the qubit electrons with spin $\mathbf{S}_{i} (i=1,2)$; 
by appropriate gating one can ensure that the system dots always stay occupied with a single electron each 
which opens up the possibility for storage of spin-spin entanglement between different (remote) quantum dots. 
%\textcolor{blue}{and in this way they can store spin entanglement.} 
An electron occupying the ancilla dot $j$ interacts locally with the system spin $\mathbf{S}_{i}$ via the Heisenberg exchange interaction \citep{Kloeffel2013}
\begin{eqnarray}
H_{\text{IN}}^{i,j} & = & J_{i,j} \mathbf{S}_{i} \cdot \boldsymbol{\sigma}_{j}, \label{eq:heisenberg}
\end{eqnarray}
where $\boldsymbol{\sigma}_{j} = \frac{1}{2} \sum_{\sigma,\sigma'}d_{j\sigma}^{\dagger} \boldsymbol{\tau}_{\sigma,\sigma'} d_{j\sigma'}$ refers to the spin-$1/2$ ancilla operator; here, $d_{j\sigma}^{\dagger}$ creates an electron with spin $\sigma={{\uparrow}},{{\downarrow}}$ in the ancilla dot $j$ and $ \boldsymbol{\tau}$ is the vector of Pauli matrices.
The exchange coupling $J_{i,j}$ can be as large as several tens of $\mu\mathrm{eV}$ and controlled \textit{in situ} by gating of the tunneling barrier between 
two nearby dots  \citep{HKP+07, Kloeffel2013}.

The system is subject to an external magnetic field $\mathbf{B}$, taken along $\hat{z}$. 
In a suitable rotating frame the global homogeneous magnetic field drops out from the dynamics, and we are left with (small) inhomogeneous gradient fields, described by the Zeeman Hamiltonian 
\begin{eqnarray}
H_{\text{Z}} & = & \sum_{i}\delta_{i}S_{i}^{z}. \label{eq:ZeemanH}
\end{eqnarray}
Here, the magnetic gradients $\delta_{i} {\lesssim} 2\mu\mathrm{eV}$ can be engineered via
on-probe 
micro- \citep{PLOT+08} or nanomagnets \citep{Forster2015}
and/or nuclear Overhauser fields \citep{Kloeffel2013}.

\begin{figure}
\includegraphics[width=0.9\columnwidth]{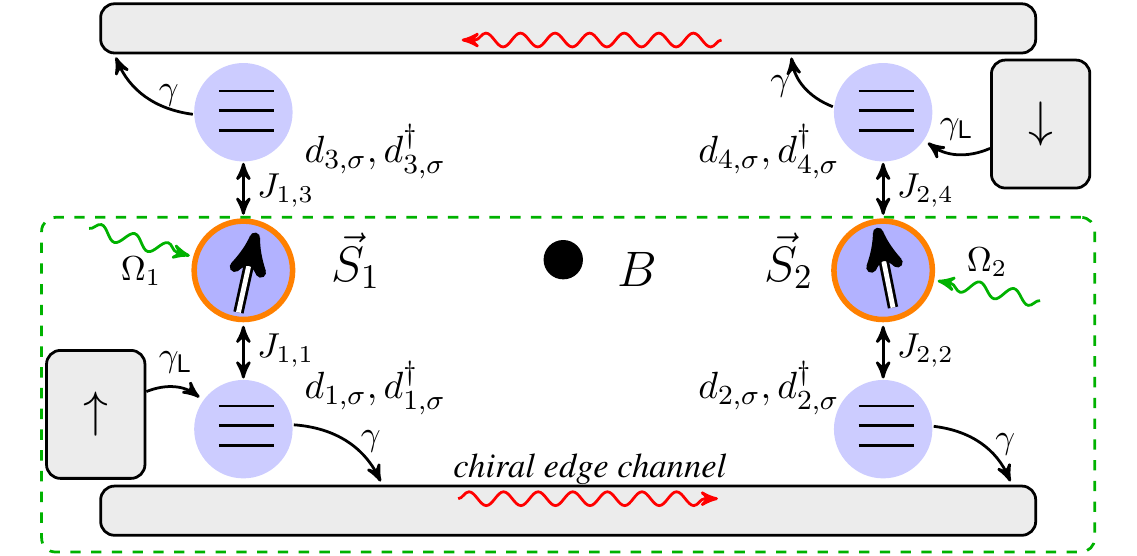}

\protect\caption{\label{fig:setup1}(color online). Scheme of the QHE-based setups. 
Two spatially separated qubits ($\mathbf{S}_{1},\mathbf{S}_{2}$) are coupled to auxiliary QDs, which are interconnected by a unidirectional QH edge channel. 
The upstream ancilla dot(s) are pumped selectively from a Fermi reservoir with a rate $\gamma_{\text{L}}$. 
While the first (purely dissipative) scheme requires two separate QHE channels, for the second scheme a single channel suffices (dashed box) together with local ESR driving fields of strength $\Omega_{i}$.}
\end{figure}

%-----------------------------------------------------------------------------
\subsection{Transport via QHE states}
\label{subsec:QHE}
A two-dimensional electron gas (2DEG) in a large magnetic field supports QHE channels which 
have proven to
provide an ideal test bed for electronic-optics-like experiments, since they allow for ballistic, one-dimensional, and chiral electron transport  \citep{Bocquillon2014};
with backscattering drastically reduced due to chirality, in the QH regime the mean-free path of electrons is increased up to $\sim (0.1-1) \mathrm{mm}$ \citep{Komiyama1992, Ji2003, Stace2004}. 
Let us consider two nodes, consisting of 
just 
one system and one ancilla dot each, with the ancilla dots interconnected by such 
a chiral edge channel; 
compare the dashed box in Fig.~\ref{fig:setup1}. 
To describe the dynamical evolution of the system and ancilla degrees of freedom of this cascaded quantum system \citep{Stannigel2012,Gardiner2004}, we trace out the channel and employ the fermionic input-output formalism (see Appendix~\ref{sec:cascaded-meq}) \citep{Stace2004, Gardiner2004}. 
We then arrive at the following Markovian ME for the reduced DM of system and ancilla dots, 
\begin{eqnarray}
\dot{\varrho} =  -i\left[H_{\text{Z}}+H_{\text{IN}},\varrho\right]+{\cal L}_{\text{tr}}\varrho, \label{eq:MEtotal}
\end{eqnarray} 
where $H_{\text{Z}}$ accounts for Zeeman energies 
[compare Eq.~(\ref{eq:ZeemanH})], 
$H_{\text{IN}}$  describes \textit{local} spin-spin interactions between system and auxiliary dots
\begin{eqnarray}
H_{\text{IN}} = \sum_{\left<i,j\right>} H_{\text{IN}}^{i,j}, \label{eq:HIN}
\end{eqnarray} 
and ${\cal L}_{\text{tr}}\varrho=\sum_{\sigma} {\cal L}_{\text{tr},\sigma}\varrho$ describes electron transport. 
The latter reads explicitly 
\begin{eqnarray}
{\cal L}_{\text{tr},\sigma}\varrho & = & \frac{\gamma_{\text{L},\sigma}}{2}{\cal D}\left[d_{1\sigma}^{\dagger}\right]\varrho+\frac{\gamma}{2}{\cal D}\left[d_{1\sigma}+d_{2\sigma}\right]\varrho\nonumber \\
 & + & {\frac{\gamma}{2}}\left[d_{1\sigma}^{\dagger}d_{2\sigma}-d_{2\sigma}^{\dagger}d_{1\sigma},\varrho\right]. \label{eq:cascaded}
\end{eqnarray}  
Here, the first term describes spin-selective pumping of the first ancilla dot, which could be achieved either via ferromagnetic leads or spin-filtering techniques \citep{Hanson2004}; 
in our dissipative setup, electron pumping (resulting in an effective electron source) is required in order to obtain a genuine nonequilibrium situation with continuous electron driving.
The last two terms give the \textit{nonlocal} incoherent and coherent contributions of the channel-mediated coupling between the ancilla dots, respectively.  
The theoretical treatment 
underlying Eq.~(\ref{eq:cascaded}) assumes
weak coupling to the reservoir and  a flat reservoir spectral density (Born-Markov approximation),
an idealized dispersion-free channel, and the spin-resolved ancilla dot levels to be aligned within ${\lesssim} \gamma$ \citep{Stace2004}.
Lastly, in accordance with the cascaded nature of the system, $\varrho$ in Eq.~(\ref{eq:cascaded}) accounts for a time delay between the nodes.
For distances $\sim \mu\mathrm{m}$, however, one can neglect this time delay, since electron transport happens  quasi-instantaneously on the relevant time scales 
(see Appendix~\ref{sec:ancilla} for an extended discussion).

For fast dissipation $(\gamma, \gamma_{\mathrm{L}} {\gg} J)$, the auxiliary dots settle into a quasisteady state  ($\rho_{\text{a}}^{\text{ss}}$) on a time scale much shorter than the relevant system-dots dynamics. 
In this case, 
the system-bath coupling $H_{\text{IN}}$ can be treated perturbatively and 
one can adiabatically eliminate the ancilla coordinates yielding a coarse-grained equation of motion for the system spins ($\mathbf{S}_{1},\mathbf{S}_{2}$). 
The subsequent full calculation follows the general framework developed in \cite{Kessler2012} and
is presented in detail in Appendix~\ref{sec:Adiabatic-Elimination-and-effective-equation}.
%Ref.~\citep{supplement}.
The ensuing first-order contributions $\sim J$ result in effective, local magnetic fields for the system spins $\mathbf{S}_{i}$, 
which are oriented along the quantization axis $z$ and 
given by the mean value of the ancilla spins in the quasisteady state; i.e., 
$\left\langle \sigma_{i}^{z}\right\rangle _{\text{ss}}=\text{tr}_{\text{a}}\left\{ \sigma_{i}^{z}\rho_{\text{a}}^{\text{ss}}\right\} $ ($\mathrm{tr_{a}}[\dots]$ 
denotes the trace over the auxiliary degrees of freedom).
As discussed in more detail below, via a suitable choice of local magnetic gradients $\delta_{i}$ in Eq.~(\ref{eq:ZeemanH}) these first-order terms can be chosen to vanish. 
To second order,
nonlocal charge correlations inherent to the ancilla system are transferred to the system spins resulting in an effective master equation with one dominant nonlocal term. 
It reads $\Gamma_{+}^{\text{ff}}{\cal D}[{\bf v}_{\text{ff}}^{+}\cdot\left(S_{1}^{+},S_{2}^{+}\right)]\rho$,
where $\rho=\mathrm{tr_{a}}[\varrho]$  
%where $\mathrm{tr_{a}}[\dots]$ denotes the trace over the auxiliary degrees of freedom, 
and 
${\bf v}_{\text{ff}}^{+}=\left(\cos\frac{\theta_{\text{ff}}}{2},\sin\frac{\theta_{\text{ff}}}{2}\right)$. 
Explicit expressions for $\theta_{\text{ff}}$ and $\Gamma_{+}^{\text{ff}}$ 
can be found in Appendix~\ref{sec:Effective-master-equation}. 
This nonlocal Lindblad term features two stationary states: 
$\left|\Psi_{\text{ss},1}\right>=\cos\frac{\theta_{\text{ff}}}{2} \left| \uparrow\downarrow \right>-\sin\frac{\theta_{\text{ff}}}{2} \left|\downarrow\uparrow\right>$ and a simple product state 
$\left|\Psi_{\text{ss},2}\right>= \left| \uparrow\uparrow\right>$.
To destabilize the second (unentangled) stationary solution,
we can either (i) add an extra channel or (ii) apply a coherent driving to the localized spins in order to (approximately) recover the dynamics stated in Eqs.~(\ref{eq:goal1}) and ~(\ref{eq:goal2}), respectively.
In this scenario (as opposed to the situation with just one nonlocal Lindblad term), the steady state is unique, which makes the scheme robust against initialization errors.

\subsubsection{Two channels and no driving}
To mimic Eq.~(\ref{eq:goal1}), we consider a purely dissipative setting with two separate edge channels that are pumped spin-selectively by spin-up (spin-down) electrons only, respectively, interacting through different ancilla dots with the qubits; compare Fig.~\ref{fig:setup1}. 
Here, two separate channels are introduced in order to effectively obtain not only one, but two independent, nonlocal jump operators. 
The latter is needed to (approximately) emulate the paradigm master equation (1) with two independent jump operators, which (under the conditions specified in Sec.~\ref{sec:dissip}) ensures a unique steady state. 
The spin of the injected electron determines the type of nonlocal jump operator in the effective master equation for the system spins: Injecting a spin-up electron into the ancilla system will result in a collective flip $\mathcal{D}[\mu S_{1}^{+} + \nu S_{2}^{+}]\rho$, because the ancilla electron can only flip to spin-down (which comes with a spin-raising flip to the system spins), whereas injecting a spin-down electron into the ancilla system will lead to a collective flip of the form $\mathcal{D}[\nu S_{1}^{-} + \mu S_{2}^{-}]\rho$, because the ancilla electron can only flip to spin-up (which comes with a spin-lowering flip to the system spins). 
In this setting, the quantized levels in the ancilla dots help to suppress undesired, parasitic local processes where electrons are transferred from the lower (upper) to the upper (lower) edge channel by virtually occupying the system dot. 
For $J_{1}\equiv J_{1,1}=J_{2,4}$ and $J_{2}\equiv J_{2,2}=J_{1,3}$, the ensuing effective ME for the two qubits only reads  
\begin{eqnarray}
\dot{\rho} & = & +\Gamma_{+}^{\text{ff}}{\cal D}\left[{\bf v}_{\text{ff}}^{+}\cdot\left(S_{1}^{+},S_{2}^{+}\right)\right]\rho\label{eq:proposal1}\\
 &  & +\Gamma_{+}^{\text{ff}}{\cal D}\left[{\bf v}_{\text{ff}}^{+}\cdot\left(S_{2}^{-},S_{1}^{-}\right)\right]\rho+{\cal L}_{\text{n-id}}^{(1)}\rho. \nonumber 
\end{eqnarray}
Here, the external magnetic gradients have been chosen as $\delta_{1(2)}=\mp\left(J_{1}\langle\sigma_{1}^{z}\rangle_{ss}-J_{2}\langle\sigma_{2}^{z}\rangle_{ss}\right)$ (the index in parentheses refers to the lower sign) in order to cancel the first-order terms $\sim J$. 
Realistic numerical values for $\delta_{i}$ will be provided below.
%Here, $\rho{=}\mathrm{tr_{a}}[\varrho]$, where $\mathrm{tr_{a}}[\dots]$ denotes the trace over the auxiliary degrees of freedom, and ${\bf v}_{\text{ff}}^{+}{=}\left(\cos\frac{\theta_{\text{ff}}}{2},\sin\frac{\theta_{\text{ff}}}{2}\right)$. 
Explicit expressions for 
the mixing angle $\theta_{\text{ff}}$, 
the effective (second-order $\sim J^2$) rate $\Gamma_{+}^{\text{ff}}$ and the undesired terms ${\cal L}_{\text{n-id}}^{(1)}$
can be found in Appendix~\ref{sec:Effective-master-equation}. 
The ME given in Eq.~(\ref{eq:proposal1}) 
indeed features \textit{nonlocal} transport-mediated jump terms of the same 
squeezing-type
form as given in Eq.~(\ref{eq:goal1}), with $\mu\equiv\cos\frac{\theta_{\text{ff}}}{2}$ and $\nu\equiv\sin\frac{\theta_{\text{ff}}}{2}$; see inset in Fig.~\ref{fig:result1}a). 

%\textcolor{blue}{Note that if the coupling between the system spins and the quantum channel was direct, without ancilla QDs, it would be impossible to avoid undesired processes %where electrons are transferred from the lower (upper) to the upper (lower) channel by virtually occupying the system dot.}

\subsubsection{One channel and driving}
Next, we follow the same strategy to (approximately) recover Eq.~(\ref{eq:goal2}).
To do so, we consider a potentially simpler setup, where a single channel suffices, but an additional (weak) resonant drive needs to be introduced; compare Fig.~\ref{fig:setup1}. 
As shown in detail in Appendix~\ref{sec:Effective-master-equation}, again for $\gamma, \gamma_{\mathrm{L}} {\gg} J$, this system is
described by 
\begin{eqnarray}
\dot{\rho} & = & -i\left[H_{\text{d}},\rho\right]-\Delta\left[S_{2}^{-}S_{1}^{+}-S_{1}^{-}S_{2}^{+},\rho\right]\label{eq:proposal2}\\
 & + & \Gamma_{+}^{\text{ff}}{\cal D}[{\bf v}_{\text{ff}}^{+}\cdot(S_{1}^{+},S_{2}^{+})]\rho+{\cal L}_{\text{n-id}}^{(2)}\rho,\nonumber 
\end{eqnarray} 
where $H_{\text{d}}=\sum_{i=1,2}2\Omega_{i}S_{i}^{x}$ describes electron-spin-resonance (ESR) driving of the spins in the rotating frame, and $\Delta$ is an effective, 
coherent spin-spin interaction
mediated by the channel. 
Explicit expressions for $\theta_{\text{ff}}$, $\Gamma_{+}^{\text{ff}}$, $\Delta$ and ${\cal L}_{\text{n-id}}^{(2)}$ 
can be found in Appendix~\ref{sec:Effective-master-equation}. 
Here, the Zeeman energies have been chosen as $\delta_{i}=-J_{i}\langle\sigma_{i}^{z}\rangle_{ss}$.
Again, realistic numerical values for $\delta_{i}$ will be provided below.

As evident from Eqs.~(\ref{eq:proposal1}) and (\ref{eq:proposal2}) the \textit{continuous} interaction of
the two spin qubits with the entangled steady state of the ancilla electrons gives
rise to  more than just the desired Lindblad terms; cf. also Fig.~\ref{fig:rates}.% in Appendix~\ref{sec:Adiabatic-Elimination-and-effective-equation}.   
To address this limitation, we discuss below an alternative \textit{stroboscopic} (that is, not continuous) setup
which allows for better control of the system-ancilla interactions and
therefore yields more ideal effective dynamics (as discussed in Sec.~\ref{sec:dissip}).

\begin{figure}
\includegraphics[width=0.9\columnwidth]{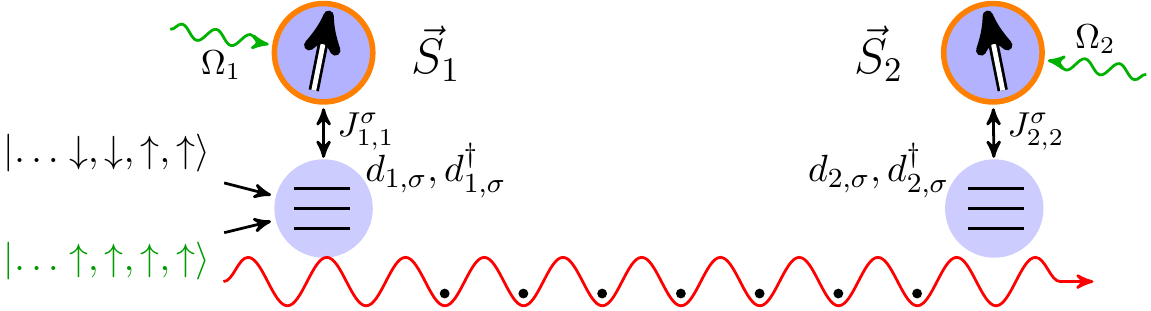}

\protect\caption{\label{fig:setup2}(color online). Scheme of the SAW-based setups. 
Two spatially separated qubits ($\mathbf{S}_{1},\mathbf{S}_{2}$) are coupled to auxiliary QDs, which are interconnected by a depleted one-dimensional channel. 
Via 
mobile dots single electrons are continuously transferred between the two ancilla dots, where they interact successively with the system spins $\mathbf{S}_{i}$ for a controlled interaction time $\tau_{i}$.
}
\end{figure}

%-----------------------------------------------------------------------------
\subsection{Transport via SAW moving dots}
\label{subsec:SAW}
To this end we replace the edge channels by 
mobile quantum dots
based on SAWs. 
Here, we consider two ancilla QDs which are interconnected by a long depleted one-dimensional channel in a 2DEG; 
compare Fig.~\ref{fig:setup2}. 
Recently, it has been demonstrated experimentally that in such a setup SAWs can transfer reliably and on-demand single electrons from one dot to the other for distances of several micrometers \citep{Hermelin2011, McNeil2011}, with the potential to extend this to hundreds of micrometers \citep{Bertrand}. 
Our protocol then consists of a continuous train of mobile dots that interact successively with the two system spins $\mathbf{S}_{i}$ for a (electrostatically) controlled time $\tau_{i}$, very much like in a conveyor belt. 
Therefore, for a single ancilla electron the protocol comprises five steps: (i) load the first ancilla dot with electron spin $\sigma$, (ii) interact with system spin $\mathbf{S}_{1}$ via Heisenberg coupling 
(\ref{eq:heisenberg}) for a time $\tau_{1}$, (iii) transfer the electron to the second ancilla dot (generically, $\mathbf{S}_{1}$ and the mobile electron are entangled by now), (iv) interact with system spin ${\mathbf{S}}_{2}$ via Heisenberg coupling 
(\ref{eq:heisenberg}) for a time $\tau_{2}$, and (v) eject the electron from the second ancilla dot. 
The corresponding concatenated evolution  
for the two localized spins $\mathbf{S}_{i}(i=1,2)$
can be described by \citep{Christ2007}
\begin{eqnarray}
\rho^{(n)} = \text{tr}_{\text{a}}[ e^{{\cal L}_{2,n}\tau_{2}}e^{{\cal L}_{1,n}\tau_{1}}(\rho^{(n-1)}\otimes|\sigma_{n-1}\rangle\langle\sigma_{n-1}|)], \label{eq:stroboscopicME}
\end{eqnarray}
where $\rho^{(n)}$ defines the state 
%of the system 
after the $n-$th cycle of the protocol. 
Here, the trace is taken over the ancilla degrees of freedom and the Liouvillian ${\cal L}_{i,n}$ encodes both the interaction of the auxiliary electron with the main qubit $i=1,2$
via Eq.~(\ref{eq:heisenberg}) and Zeeman terms, Eq.~(\ref{eq:ZeemanH}).  
This model assumes perfect spin transfer which is approximately correct for distances much shorter than the characteristic dephasing length scale  
which we estimate as $\sim v_{s}T_{2}^{*} {\gtrsim} 100\mu\mathrm{m}$
for $v_{s}{\approx} 3\mu\mathrm{m}/\mathrm{ns}$ and $T_{2}^{*}{\approx} 100\mathrm{ns}$ \citep{McNeil2011}. 

Along the lines of our previous analysis, in what follows we present two SAW-based schemes: (i) a protocol with alternating spin directions and suitably synchronized exchange couplings and (ii) a spin-polarized protocol with a coherent driving. Both transport protocols will be shown to drive the localized spins to an entangled steady state, independently of the initial state. 

\subsubsection{Alternating spin sequences}
To recover the purely dissipative dynamics (\ref{eq:goal1}), we assume alternating spin sequences (as could be realized by proper spin filtering on subnanosecond time scales \citep{Hanson2004}), together with appropriately synchronized interaction times $\tau_{i}$ or exchange couplings $J_{i}^{\sigma}\equiv J_{i,i}^{\sigma}$ (see Appendix~\ref{sec:app-Effective-Stroboscopic-Evolution} for a detailed derivation). 
This is necessary to achieve the desired asymmetry $\mu {\neq} \nu$. 
In the following, 
$\tau\equiv\tau_{1}=\tau_{2}$.
Then, setting $\mu = J_{1}^{{\uparrow}}\tau=J_{2}^{{\downarrow}}\tau$, $\nu = J_{2}^{{\uparrow}}\tau=J_{1}^{{\downarrow}}\tau$, up to 
${\cal O}(\tau^{3} J_{i}^{\sigma 3})$,
the evolution of the DM simplifies to
%\citep{supplement}
\begin{eqnarray}
\rho^{(n+1)} - \rho^{(n-1)} & = & \frac{1}{8} {\cal D}\left[\mu S_{1}^{+}+\nu S_{2}^{+}\right]\rho^{(n-1)}\nonumber\\
&+&\frac{1}{8} {\cal D}\left[\mu S_{2}^{-}+\nu S_{1}^{-}\right]\rho^{(n-1)},\label{eq:proposal3}
\end{eqnarray}
Here, the inhomogeneous magnetic gradients have been chosen as $\delta_{1(2)}=\mp\frac{\mu-\nu}{8\tau}$, such that all first-order terms effectively vanish.
Typical numerical values for $\delta_{i}$ will be provided below.
Indeed, we recover \textit{nonlocal} dissipators of the desired 
asymmetric (squeezing-type) form; compare Eq.~(\ref{eq:goal1}).
Alternating sequences of spin-up and spin-down electrons (with suitably synchronized couplings) then yield approximately the desired entangling dynamics.

\subsubsection{Single spin-component and driving}
Next, to emulate dynamics similar to Eq.~(\ref{eq:goal2}), we assume mobile dots with a single spin-filtered spin-component \citep{Hanson2004} and introduce an additional coherent external driving field.
In this case, for asymmetric, but time-\textit{independent} couplings ($\mu=J_{1}^{{\uparrow}}\tau$, $\nu=J_{2}^{{\uparrow}}\tau$), magnetic gradients $\delta_{i}=-J_{i}^{{\uparrow}}/4$ and weak driving $\Omega_{1,2}{\ll} J$, the evolution of the DM 
is approximately given by (see Appendix~\ref{sec:app-Effective-Stroboscopic-Evolution})
\begin{eqnarray}
\rho^{(n)} & = & \rho^{(n-1)}+\frac{\mu\nu}{8}\left[S_{1}^{-}S_{2}^{+}-S_{2}^{-}S_{1}^{+},\rho^{(n-1)}\right] \label{eq:proposal4} \\
 & + &  1/8 {\cal D}\left[\mu S_{1}^{+}+\nu S_{2}^{+}\right]\rho^{(n-1)} -2i\tau\left[H_{\text{d}},\rho^{(n-1)}\right]. \nonumber
\end{eqnarray}

Thus, we can realize the dynamics of Eqs.~(\ref{eq:goal1}) and (\ref{eq:goal2}) with arbitrary accuracy by reducing the dwell times $\tau_i$.

%-----------------------------------------------------------------------------
\section{Results and discussion}
\label{sec:results}
In the previous section, we have derived master-equation-based models for four different physical setups in total, two of them based on QHE channels and the remaining two based on SAW-induced moving quantum dots. 
In this section, we specify the experimental requirements and discuss in detail the results of our analysis, as quantified via the amount of entanglement that the different setups are able to generate between two remote spin qubits under realistic conditions. 
First, we discuss the QHE states based proposals, then the SAW-based proposals; we conclude the discussion with a comprehensive comparison of the different proposed setups.

\begin{figure}
\includegraphics[width=1\columnwidth]{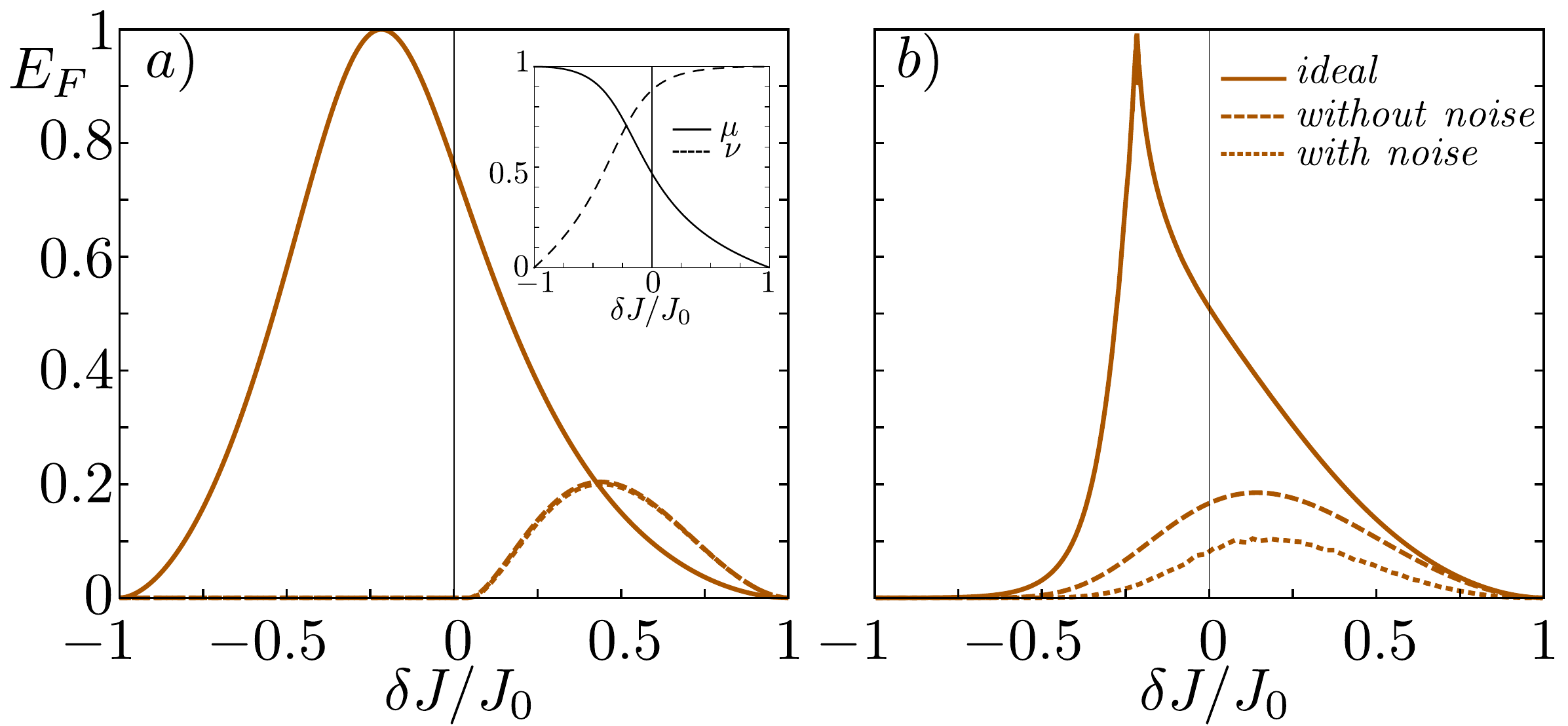}

\protect\caption{\label{fig:result1}Steady-state entanglement 
quantified via the $E_{F}$ for the two QHE-based proposals as a function of
$\delta J$. (a) and (b) are based on Eq.~(\ref{eq:proposal1}) and Eq.~(\ref{eq:proposal2}), respectively. 
The solid lines refer to the ideal result, where the peak is reached for $\mu=\nu$ (see inset). 
The dashed lines also take into account the undesired terms, described by ${\cal L}_{\text{n-id}}^{(i)}$, 
while
the dotted lines in addition account for
nuclear dephasing (see text).
Numerical parameters: $\gamma_{\text{L}}=\gamma=30\mu\text{eV}$, $J_{0}=3\mu\text{eV}$
and $\delta_{i} {\in}\left(-2,2\right)\mu\text{eV}$.
In (b), for each value of $\delta J$, 
$\Omega_{i}$ has been optimized in the range $\Omega_{i}{\in}\left(0-50\right)\text{neV}$.
}
\end{figure}

%-----------------------------------------------------------------------------
\subsection{QHE states}
\label{sec:resultsQHE}
Both Eqs.~(\ref{eq:proposal1}) and (\ref{eq:proposal2}) potentially recover 
the ideal entanglement-generating dynamics given in Eq.~(\ref{eq:goal1}) and (\ref{eq:goal2}), respectively, up to undesired terms absorbed into ${\cal L}_{\text{n-id}}^{(i)}$. 
We now turn to the central question of whether the entanglement inherent to the ideal dynamics can prevail in a realistic scenario. 
Due to the presence of the nonideal terms, even without further decoherence mechanisms, the steady state of Eqs.~(\ref{eq:proposal1}) and (\ref{eq:proposal2}) is mixed. 
We confirm and quantify its entanglement using the entanglement of formation $E_F$ (see Appendix~\ref{sec:EOF}) \cite{Wootters1998}.
As shown in Fig.~\ref{fig:result1}, for a broad range of coupling parameters ($J_{1(2)}=J_0\mp \delta J$) the generation of steady-state entanglement persists in the two schemes even in the presence of the undesired terms ${\cal L}_{\text{n-id}}^{(i)}$.
%This finding can be traced back to the spectral gap
%\citep{supplement} which indicates on what timescale the steady state is reached. 
%For optimized, but realistic parameters ($\gamma_\mathrm{L}{=}\gamma{=}10J_0{=}30\mu \text{eV}$), the gap is found to be ${\sim} 0.5 J_0^2/\gamma$ ($0.1 J_0^2/\gamma$) for the purely dissipative setting (driving scheme), which yields  ${\sim}5$ns (${\sim} 20$ns).
%Note that the steady state will not be affected significantly by noise processes slow compared to that timescale. 

\begin{figure}
\includegraphics[width=1\columnwidth]{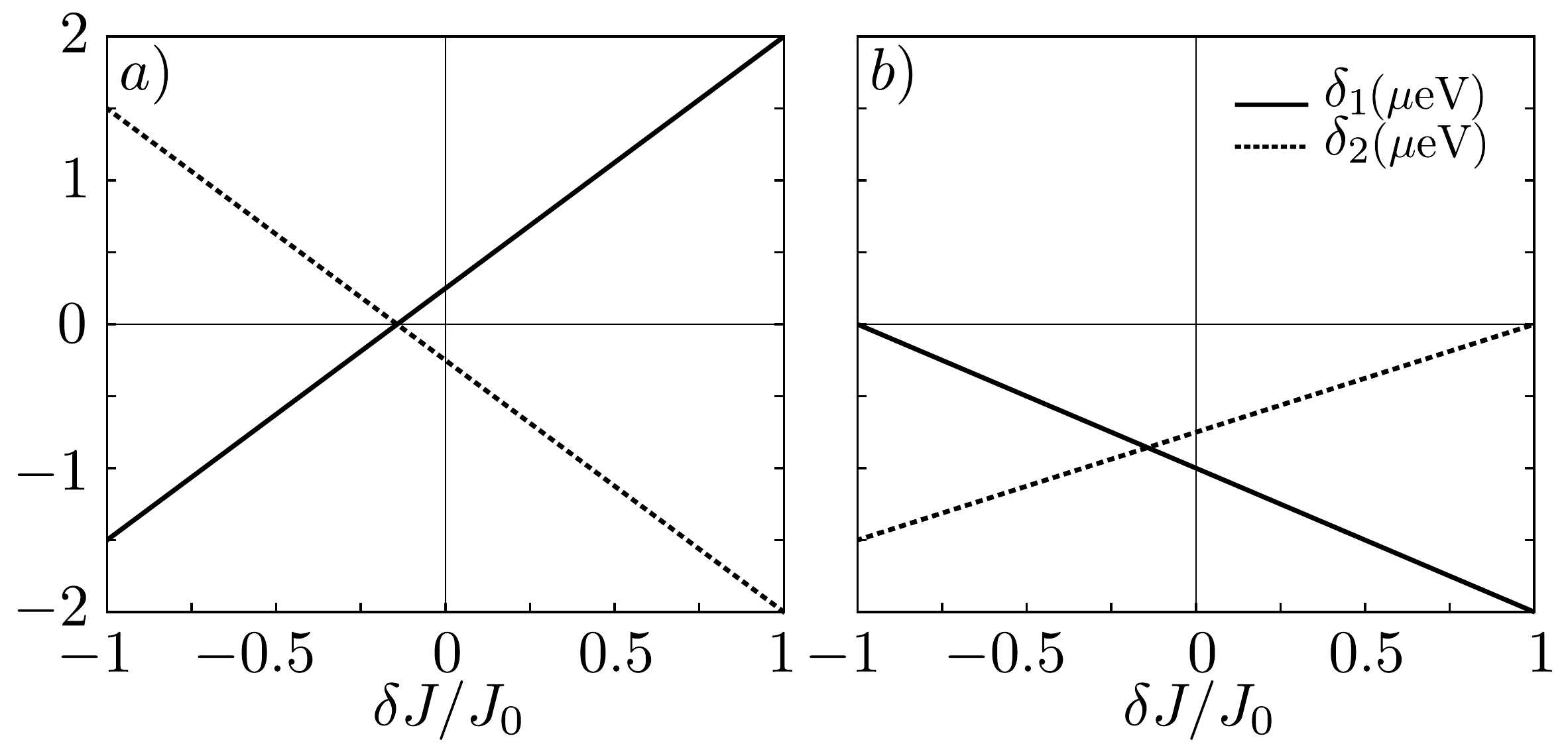}

\protect\caption{\label{fig:gradients}Value of the local magnetic fields $\delta_{1(2)}$ required to get (a) Eq.~(\ref{eq:proposal1}) and (b) Eq.~(\ref{eq:proposal2}), respectively,  as a function of $\delta J$.
%Proposal a) with two QHE channels and b) with
%coherent driving.  
Note that in (b) $\delta_{1,2}<0$ because we arbitrarily choose pumping with spin-up ancilla electrons. Correspondingly, for spin-down pumping the sign would be reversed. Numerical parameters: $\gamma_{\text{L}}=\gamma=30\mu\text{eV}$, $J_{0}=3\mu\text{eV}$.}
\end{figure}

In order to obtain sizable steady-state entanglement (which arises from nonlocal second-order effects $\sim J^2$), the first-order contributions $\sim J$ have to be canceled via local magnetic fields as described by Eq.~(\ref{eq:ZeemanH}); compare our discussion in Sec.~\ref{sec:model}. 
For $\gamma_{\text{L}}=\gamma$ (as considered in the text), the 
Zeeman energies $\delta_i$ are typically of the order of (or smaller than) the Heisenberg coupling strengths $J_i$ (i.e., typically a few $\mu$eV); see Fig.~\ref{fig:gradients}. 
Using for example nanomagnets, gradients of this size can be readily achieved  (e.g., in GaAs by local magnetic fields of a few 100mT) \citep{PLOT+08, Forster2015}.

%Taken from SM

%Taken from SM
\begin{figure}
\includegraphics[width=1\columnwidth]{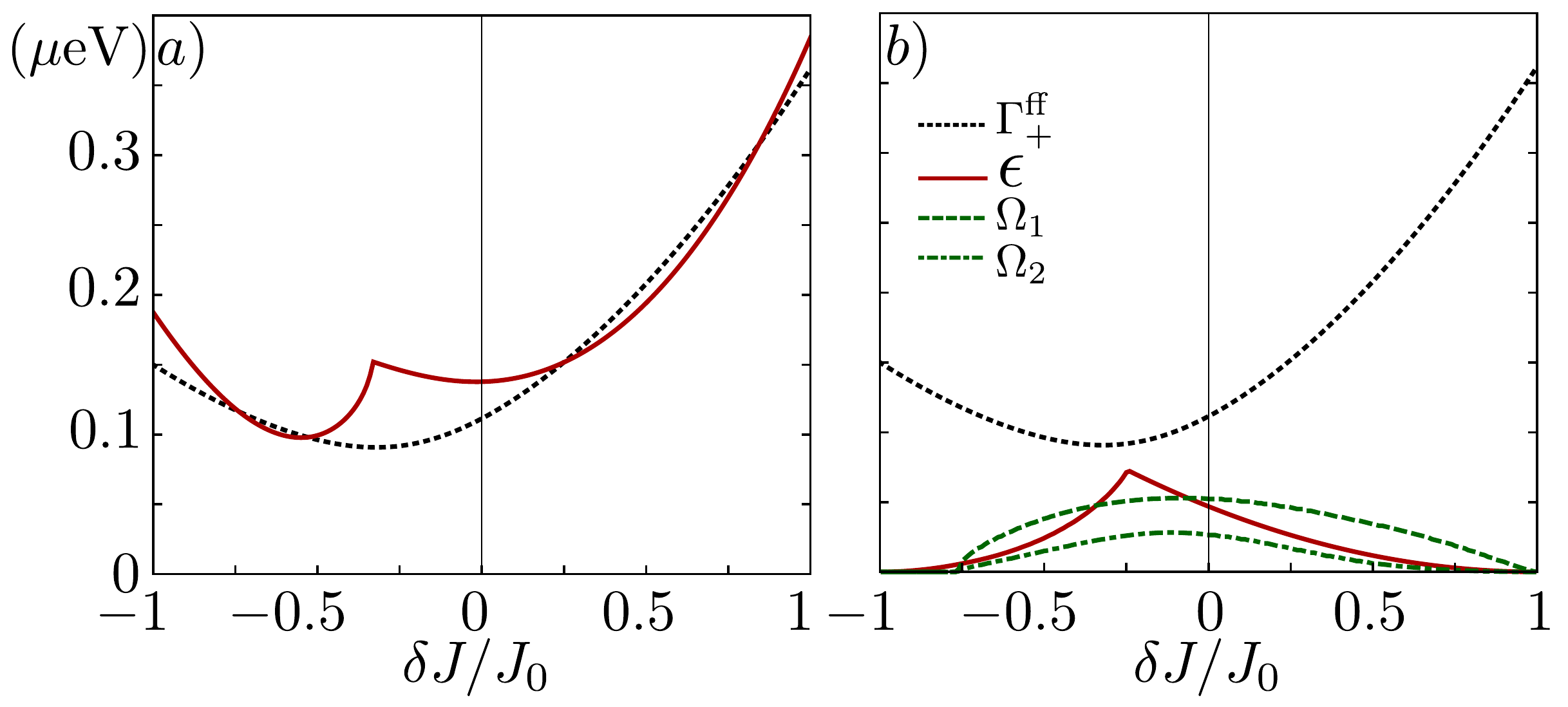}

\protect\caption{\label{fig:app-gap}(color online) Spectral gap of the dissipative dynamics (continuous
red line) and dominating rate $\Gamma_{+}^{\text{ff}}$ (dotted black line) as a function of $\delta J$.
(a) and (b) are based on Eq.~(\ref{eq:proposal1}) and Eq.~(\ref{eq:proposal2}), respectively. 
Numerical parameters: $\gamma_{\text{L}}=\gamma=30\mu\text{eV}$, $J_{0}=3\mu\text{eV}$
and $\delta_{i} {\in}\left(-2,2\right)\mu\text{eV}$.
In (b), for each value of $\delta J$, 
$\Omega_{i}$ (green lines) have been optimized in the range $\Omega_{i}{\in}\left(0-50\right)\text{neV}$.}
\end{figure}

Another important question is how long it approximately takes
for the system to reach its steady state. 
This time scale is directly related to the spectral gap of the corresponding dissipative dynamics, 
which is shown in Fig.~\ref{fig:app-gap} for the two QHE-based proposals.
%, determined by the eigenvalue of the
%Liouvillian with the largest real part different from zero, i.e.,
% $\epsilon=-\max \left\{\Re(\lambda_{i}) \right\} $,
%where $\lambda_{i}$ are the non-zero eigenvalues of the Liouvillian.
%We show this quantity in Fig.~\ref{fig:app-gap} for the two proposals.
The spectral gap is found to be proportional to $J_0^2/\gamma$, which can be increased for small values of $\gamma$, provided that the
conditions for adiabatic elimination ($J_0\ll\gamma$) are still fulfilled. 
For the parameters $\gamma=30\mu$eV and $J_0=3\mu$eV (for which the adiabatic elimination of the fast degrees of freedom is perfectly valid), 
%in the interesting parameter regime, i.e., where the maximum entanglement is generated ,
we then estimate $\epsilon\sim0.15\mu $eV and $\epsilon\sim0.03\mu $eV, respectively. 
Accordingly, the steady state is reached on a very fast time scale of roughly $\sim(5-25)$ns. 
Then, as discussed in Sec.~\ref{sec:dissip}, any noise sources or imperfections that are slow compared to this very fast, zeroth-order time scale should not affect severely the qualitative and quantitative features of the steady state. 
%These short typical times make the proposals quite robust against noise sources.

First, this is demonstrated explicitly for qubit dephasing due to nuclear spins in the (GaAs) host environment.
As explained in more detail in Appendix~\ref{sec:ap-Noise-sources}, the hyperfine interaction with the nuclei is modeled in terms of a random, slowly evolving effective magnetic field for the electron spins, yielding an extra Hamiltonian of the same form as Eq.~(\ref{eq:ZeemanH}), where the
detuning parameters $\delta_{i}$ are sampled independently from a normal distribution with standard deviation $\sigma_{\text{nuc}}$ \citep{Kloeffel2013}. 
The resulting  time-ensemble-averaged electron dephasing time $T_{2}^{*}=\sqrt{2}/\sigma_{\text{nuc}}$ has recently been extended up to $T_{2}^{*} {\approx} 3\mu\mathrm{s}$ \citep{Shulman2014}. 
As shown in Fig.~\ref{fig:result1},
already for 
$T_2^*{\approx} 30$ns, the purely dissipative scheme is basically unaffected by nuclear noise.

Second, again because of the relatively large spectral gap $\epsilon$, perfect cancellation of the first-order terms $\sim J$ is not strictly required, provided that the residual (uncanceled) magnetic fields $\Delta_{i}$ are small compared to the gap; as shown in Appendix \ref{sec:ap-Noise-sources}, typically our scheme can tolerate residual gradients $\Delta_{i}$ of up to $\sim 0.1 \mu\text{eV}$
%$\sim 150 \text{MHz}$ 
without severely affecting the generation of steady-state entanglement.

Lastly, in our analysis we have neglected several detrimental effects that may be encountered in an actual experiment, an approximation that we now justify: 
First, at sufficiently low temperatures $T{<}5\mathrm{K}$, dispersive effects and scattering out of the edge channel may be neglected for propagation distances ${\lesssim} 100\mu\mathrm{m}$ \citep{Stace2004}. 
Nevertheless, in Appendix~\ref{sec:ap-Noise-sources} we show that even a few percent of losses can be tolerated. 
Second, 
dephasing during propagation should be negligible for distances
small compared to a characteristic coherence length scale $L_{\phi}$, which we estimate as $L_{\phi} = v_{d}T_{2}^{*} {\approx} (10^{2} - 10^{3}) \mu\mathrm{m}$ for a drift velocity $v_{d} {\approx} 10^4 \mathrm{m/s}$ and (due to motional narrowing) extended dephasing time  $T_{2}^{*} {\approx} (10-100)\mathrm{ns}$ \citep{Stace2004, McNeil2011,SKG+13, Stotz2005}. 
Then, in order to suppress errors due to nonresonant dot energies, these should be controlled with a precision ${\lesssim} 1 \mu \mathrm{eV}$ \citep{Stace2004}.
Finally, based on QD experiments \citep{Hanson2004} where basically 100\% bipolar spin-filter efficiency has been demonstrated, 
we have assumed perfect spin-selective driving.
Still, with all these simplifications, the amount of steady-state entanglement  that we obtain for a realistic scenario
(with continuous ancilla-electron pumping)
is modest $(E_{F} {\approx}0.2)$ as compared to the idealized cases discussed in Eqs.~(\ref{eq:goal1}) and (\ref{eq:goal2}), respectively (even though it is still comparable to what has been predicted theoretically for two adjacent dots \citep{bohr15} and achieved experimentally for two atomic ensembles \citep{Krauter2011}). 
As shown below, one can largely circumvent this limitation by considering well-controlled stroboscopic interaction times between system and ancilla dots (as opposed to the arguably more simple continuous settings with largely fluctuating interaction times).
%Why is that so?  

%-----------------------------------------------------------------------------
\subsection{SAW moving dots}
\label{sec:resultsSAW}

\begin{figure}
\includegraphics[width=1\columnwidth]{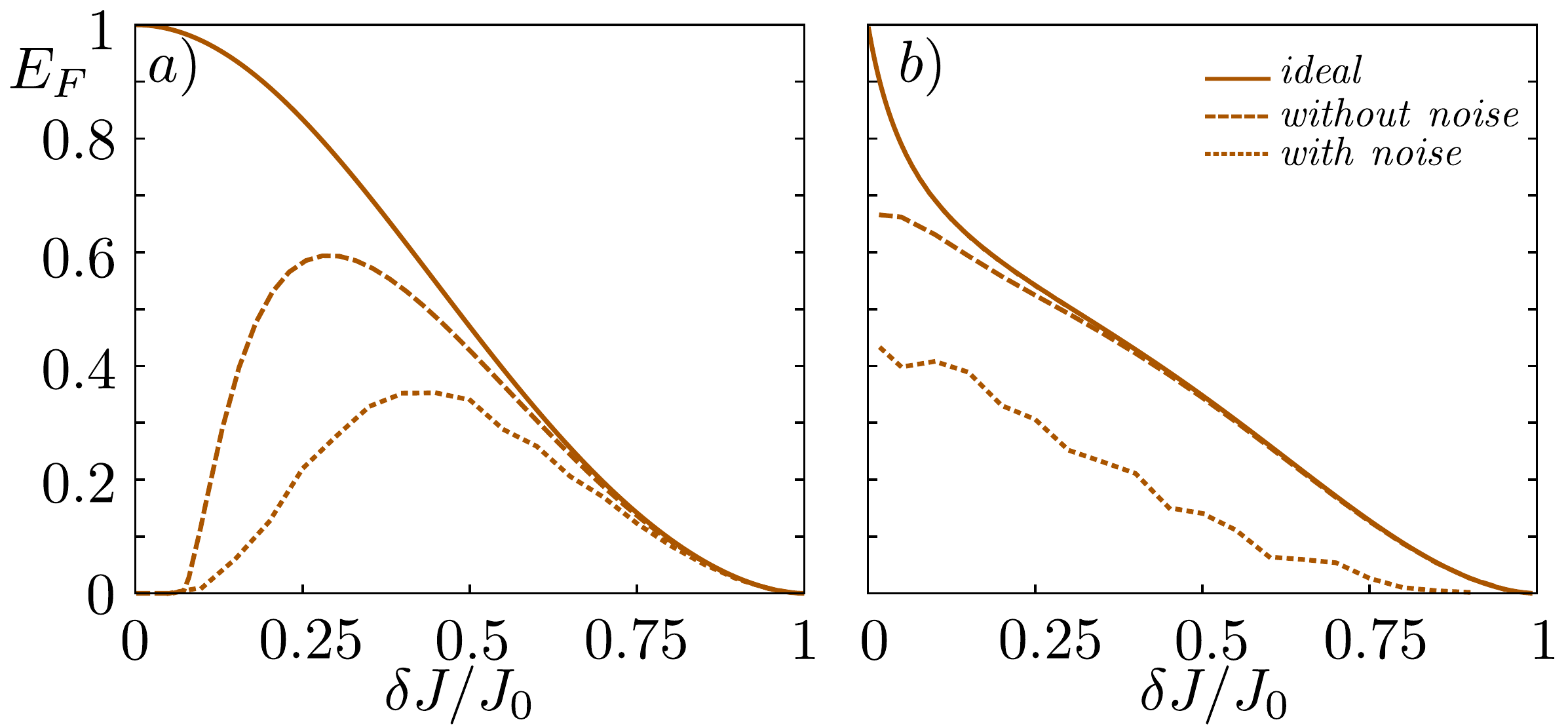}

\protect\caption{\label{fig:result2}Steady-state entanglement 
quantified via the $E_{F}$ for the two SAW-based proposals as a function of $\delta J$, with $J_{1(2)}^{{\uparrow}}=J_{0}\mp\delta J$. 
(a) and (b) are based on Eq.~(\ref{eq:proposal3}) and Eq.~(\ref{eq:proposal4}), respectively.
%a) Alternating spin sequences with synchronized couplings, Eq.~(\ref{eq:proposal3}). 
%b) Spin-filtered ancilla spins and coherent driving, Eq.~(\ref{eq:proposal4}). 
The solid lines refer to the ideal result, given by the lower order terms present in Eqs.~(\ref{eq:proposal3}) and~(\ref{eq:proposal4}), while the dashed lines correspond to the full evolution. 
The dotted lines also account for noise due to uncertainty in the dwell times and dephasing.
Numerical parameters: $\sigma_{\text{\ensuremath{\tau}}}=5\%$, $J_{0}\tau{\approx}0.38$ and
$T_{2}^{*}/\tau{\approx}300$. In (b), for each value of $\delta J$, 
$\Omega_{i}$ 
has been optimized in the range $\Omega_{i}\tau{\in}\left(0-1.5\right)\cdot 10^{-2}$.
}
\end{figure}

%Taken from SM
\begin{figure}
\includegraphics[width=1\columnwidth]{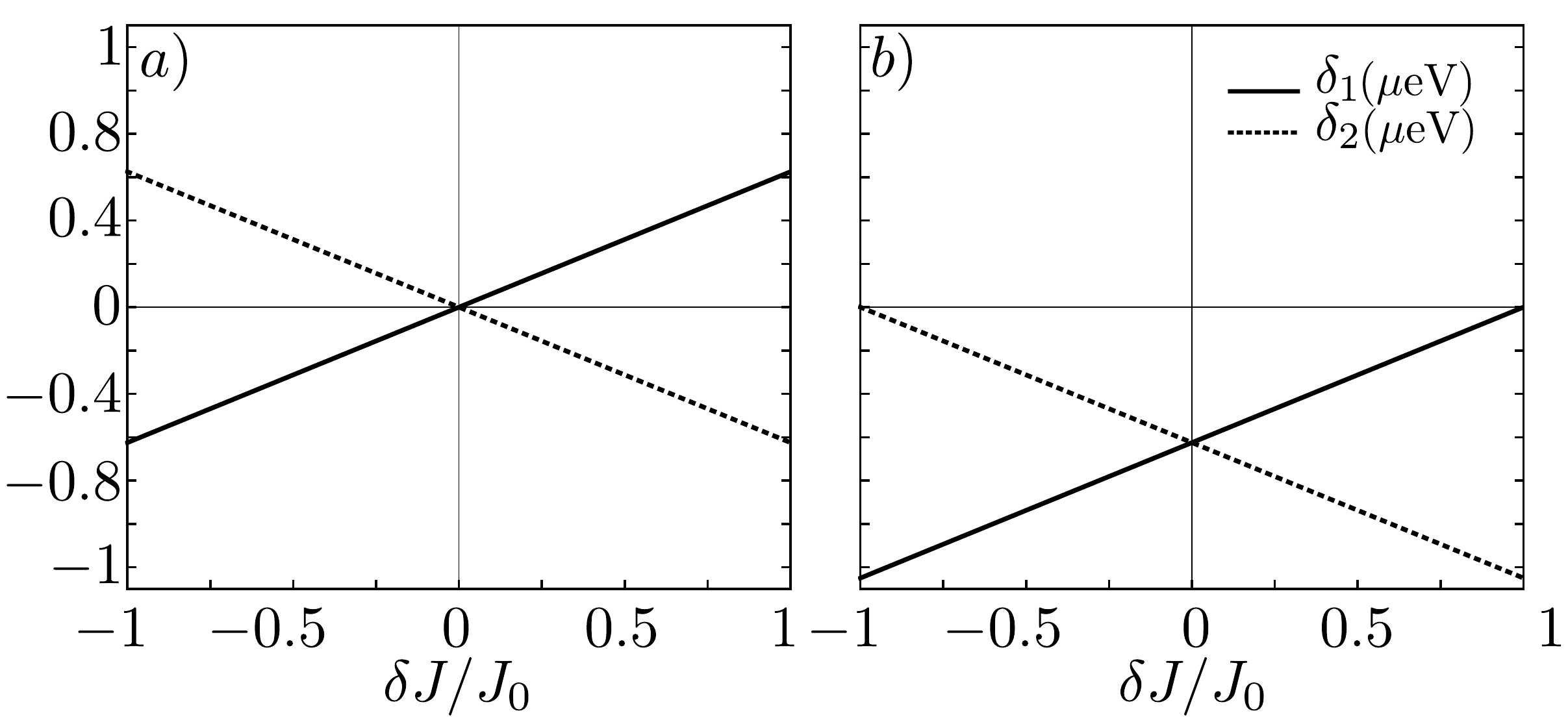}

\protect\caption{\label{fig:gradients-SAWs}Value of the local magnetic fields $\delta_{1(2)}$ required to get (a) Eq.~(\ref{eq:proposal3}) and (b) Eq.~(\ref{eq:proposal4}), respectively,  as a function of $\delta J$.
%Value of the local magnetic fields $\delta_{1(2)}$
%for the SAWs-based proposal a) with alternating spins and b) with
%coherent driving as a function of $\delta J$. 
Note that in (b) $\delta_{1,2}<0$ because we arbitrarily choose pumping with spin-up ancilla electrons. Correspondingly, for spin-down pumping the sign would be reversed. Numerical parameters: $J_{0}=2.5\mu\text{eV}$.}
\end{figure}

The dynamical equations given in 
Eqs.~(\ref{eq:proposal3}) and~(\ref{eq:proposal4})
suggest that the system qubits will be driven to an entangled steady state regardless of the initial state (as long as $\tau J_{i}\ll1$).
Our analytical results stated above have been confirmed by exact  numerical simulations of Eq.~(\ref{eq:stroboscopicME}), where the ancilla degrees of freedom have not been eliminated. 
As demonstrated in Fig.~\ref{fig:result2}, the generation of entanglement persists even in the presence of nuclear noise and residual time jitter.
%; here, the individual dwell-times $\tau_{i}$ have been chosen randomly from a normal distribution centered around $\tau$ with a standard deviation of $\sigma_{\tau}$ 
We include this noise
source by choosing the interaction times $\tau_{i}$ randomly from a Gaussian
distribution centered around the average $\tau$ with a
standard deviation of $\sigma_{\tau}$
(see Appendix~\ref{sec:ap-Noise-sources} for a detailed analysis of noise sources). 
For sufficiently low time jitter and typical dephasing times $T_2^{*}=(30-300)\mathrm{ns}$, we find 
$E_{F} {\gtrsim} 0.4$, 
which extends up to $E_{F} {\gtrsim} 0.7$ for $T_2^{*}{\approx} 1\mu\mathrm{s}$. 
Typically, the steady state is reached after $\sim 10^3$ iterations, that is, within $\sim (0.1-1)\mu \mathrm{s}$ for $\tau {\approx} (0.1-1)\mathrm{ns}$. 
The local Zeeman energies required to effectively cancel the first-order terms are shown in Fig.~\ref{fig:gradients-SAWs}. 
However, we have also checked numerically that perfect cancellation of the first-order terms is not strictly required (for details see Appendix \ref{sec:ap-Noise-sources}); accordingly, residual gradients of up to $\sim 0.03 \mu\text{eV}$ can be tolerated without severely affecting our results.

\begin{figure}
\includegraphics[width=1\columnwidth]{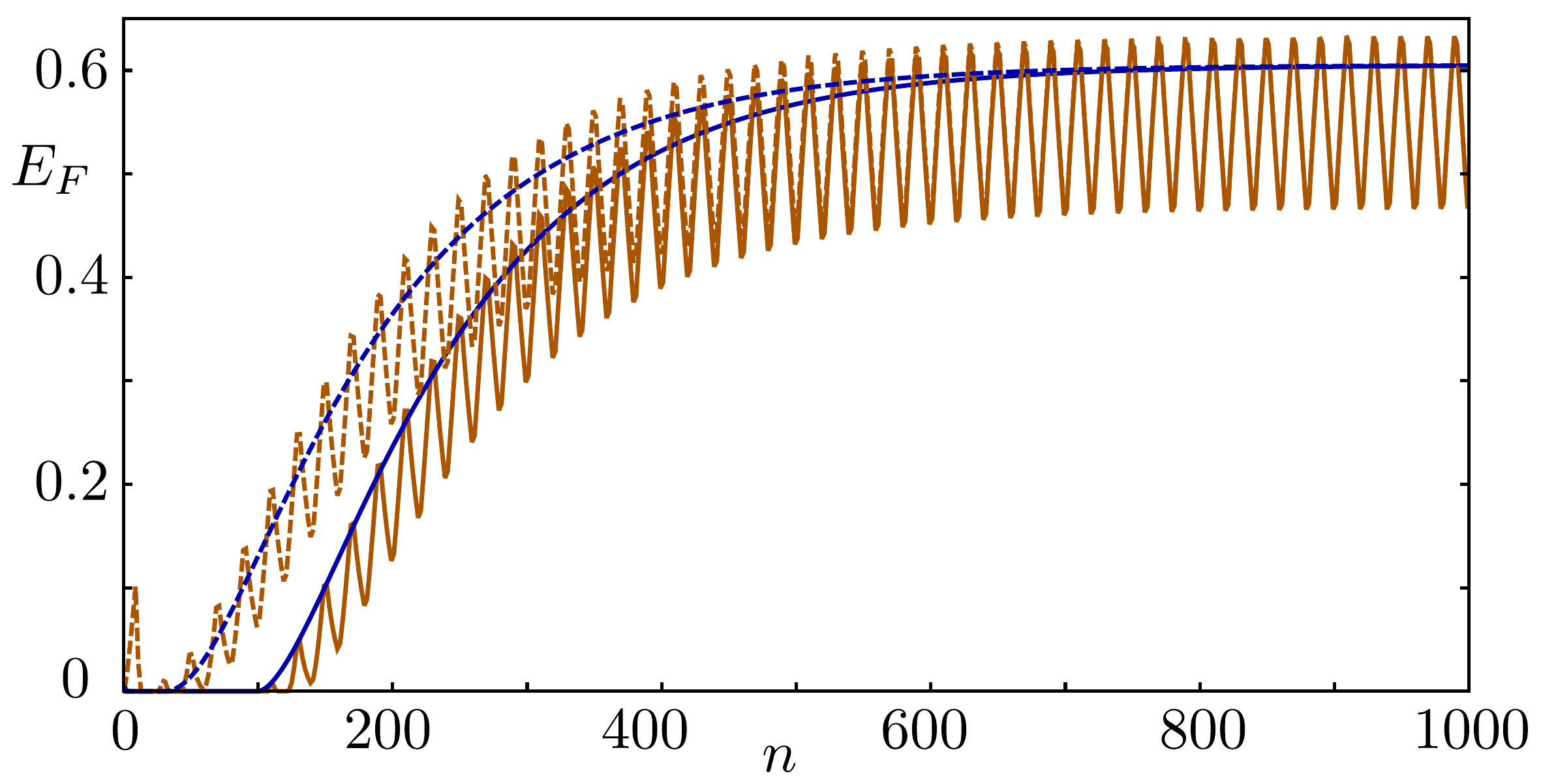}

\protect\caption{\label{fig:app-strob-Passive-scheme-with-mobiledots}(color online). Steady-state entanglement 
quantified via the $E_{F}$ for the SAW-based proposal corresponding to Eq.~(\ref{eq:proposal3})
%Entanglement between two remote qubits quantified via the $E_F$ for the  SAW-based proposal with alternating
%spin sequences and synchronized couplings 
as a function of time ($t=2\text{n}\tau$) for two different initial states (continuous and dashed lines, respectively). 
Blue: Alternating spins. Orange: Alternating sequences
of ten spins. Numerical parameters: $\delta J/J_{0}=0.28$ and 
 $J_{0}\tau{\approx}0.38$
%$J_{0}=2.5\mu eV$ 
%and $\tau=0.1\text{ns}$
.}
\end{figure}

The ideal, analytical result given in Eq.~(\ref{eq:proposal3}) assumes the injection of alternating spin components of the form $\uparrow, \downarrow, \uparrow, \dots$. However, this condition can be relaxed to longer sequences of aligned ancilla spins, of the form $\uparrow, \uparrow, \dots, \downarrow, \downarrow,\dots,\uparrow,\uparrow,\dots$. This has been confirmed numerically in Fig.~\ref{fig:app-strob-Passive-scheme-with-mobiledots}. Accordingly, the switching times of the gates can be increased by about an order of magnitude without severely affecting the amount of steady-state entanglement.  
  
%Taken from SM

\subsection{Comparison of the Setups}

The presented proposals based on QHE states constitute continuous
entangling generating setups in the sense that once the setup has been
prepared there is no need to interact externally with the system
before the entanglement measurement; 
moreover, they have been shown to drive the system
to the steady state on very fast time scales (in a matter of few ns). 
However, this (arguably simple) continuous
setting comes with the disadvantage of undesired terms in the master
equations ~(\ref{eq:proposal1}) and~(\ref{eq:proposal2}). As a
consequence, even in the cleanest setup, we cannot go beyond a
steady-state entanglement of $E_F\approx 0.2$ebits.
As evidenced by our stroboscopic SAW-based scheme, 
this limitation can be overcome by suitably controlling the electron dwell times $\tau_i$ in the ancilla dots. 
%This limitation is overcome in the SAWs based setups, where the
In this way, the effective dynamics given in Eqs.~(\ref{eq:proposal3})
and~(\ref{eq:proposal4}) can be ensured to approach the ideal ones
(by controlling the dwell times $\tau_i$). Therefore, in the limit
$\tau_i\rightarrow0$ and without noise sources, we would recover
the pure entangled steady states of Eqs.(\ref{eq:goal1}) and (\ref{eq:goal2}) and
could approach \textit{perfect} entanglement ($E_F = 1$).
%\textit{perfect} entanglement ($E_F=1$). 
Here, we estimate an upper limit of $
E_F\approx 0.7$ when accounting for typical experimental parameters and imperfections.
This better performance comes with the experimental
challenge to transport many electrons via (for example) the SAW-created
potentials reliably and with accurate (electrical) control of the electronic dwell
times. Moreover, the proposal with alternating spin sequences comes with
further requirements as the proper spin-filtering synchronized with
the exchange couplings. 
However, based on recent progress demonstrated for 
single-electron transport experiments with SAW moving dots \citep{Hermelin2011, McNeil2011, Bertrand} 
and the robustness against errors
(as we demonstrate here) 
%give some hope on the possibility to perform the given experiment.
a future, successful experimental realization of our scheme should be feasible. 

%\begin{figure}
%\includegraphics[width=1\columnwidth]{ED-saws-cut}
%
%\protect\caption{\label{fig:ED}\textcolor{blue}{(color online). Upper and lower bounds of distillable
%entanglement in the steady-state  
%quantified via the $E_{F}$ and $E_{D\to}$ for the two SAW-based proposals as a function of $\delta J$, with $J_{1(2)}^{{\uparrow}}=J_{0}\mp\delta J$. 
%a) (b)) results are based on Eq.~(\ref{eq:proposal3}) and Eq.~(\ref{eq:proposal4}), respectively.
%%a) Alternating spin sequences with synchronized couplings. 
%%b) Spin-filtered ancilla spins and coherent driving. 
%The solid and dotdashed lines correspond to the full evolution, while 
%the dashed and dotted lines account for noise due to uncertainty in the dwell times and nuclear dephasing.
%Numerical parameters: $\sigma_{\text{\ensuremath{\tau}}}=5\%$, $J_{0}\tau{\approx}0.38$ and
%$T_{2}^{*}/\tau{\approx}300$. In b), for each value of $\delta J$, 
%$\Omega_{i}$ 
%has been optimized in the range $\Omega_{i}\tau{\in}\left(0-1.5\right)\cdot 10^{-2}$.}}
%\end{figure}

%abstract for new figure
\begin{figure}
\includegraphics[width=1\columnwidth]{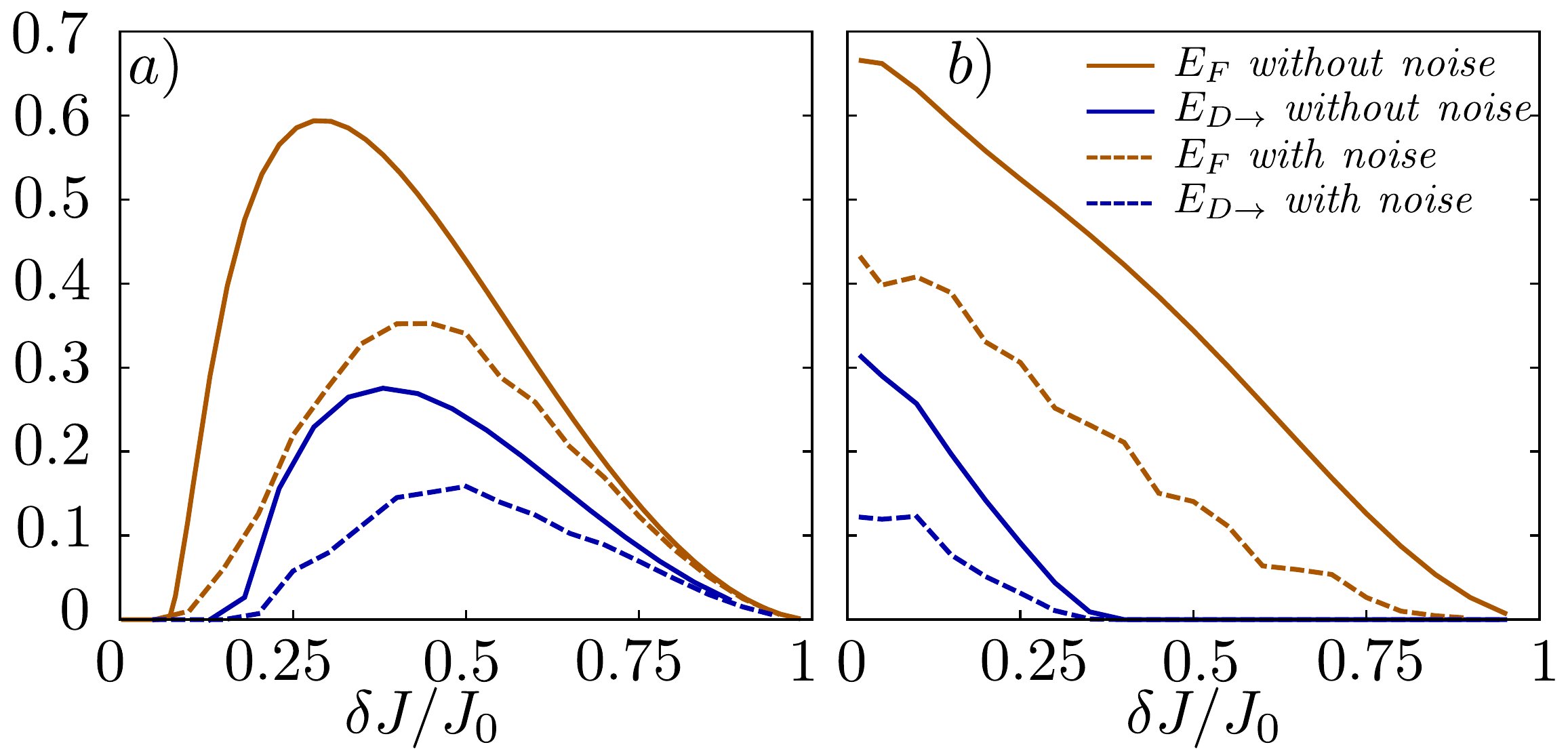}

\protect\caption{\label{fig:ED}(color online). Upper and lower bounds of distillable
entanglement in the steady-state  
quantified via the $E_{F}$ (orange) and $E_{D\to}$ (blue) for the two SAW-based proposals as a function of $\delta J$, with $J_{1(2)}^{{\uparrow}}=J_{0}\mp\delta J$. 
(a) and (b) show results based on Eq.~(\ref{eq:proposal3}) and Eq.~(\ref{eq:proposal4}), respectively.
%a) Alternating spin sequences with synchronized couplings. 
%b) Spin-filtered ancilla spins and coherent driving. 
The solid  lines correspond to the full evolution, while 
the dashed  lines account for noise due to uncertainty in the dwell times and nuclear dephasing.
Numerical parameters: $\sigma_{\text{\ensuremath{\tau}}}=5\%$, $J_{0}\tau{\approx}0.38$ and
$T_{2}^{*}/\tau{\approx}300$. In (b), for each value of $\delta J$, 
$\Omega_{i}$ 
has been optimized in the range $\Omega_{i}\tau{\in}\left(0-1.5\right)\cdot 10^{-2}$.}
\end{figure}

Given the additional experimental challenges for an accurate control of the ancilla electron dwell times $\tau_{i}$
with synchronized (electrical) control of the Heisenberg coupling constants,
one may wonder whether the increase in obtainable steady-state entanglement (in the stroboscopic SAW-based schemes) is worth the effort. 
This, of course, depends on the ultimate purpose of entanglement generation. 
When viewing entanglement production mainly as an experimental benchmark to demonstrate the capability to entangle, 
any entanglement measure (such as our canonical choice, the entanglement of formation $E_{\text{F}}$) would do;
%and we use a canonical measure (the entanglement of formation $E_{F}$)
any state with nonzero $E_{F}$ can be shown (in principle) to be entangled either by measuring a suitable entanglement witness 
or by sufficiently precise state tomography. 
However, $E_{F}$ will not tell us, in general, how useful the state is for subsequent QIP tasks.
Since most applications of
entanglement require almost pure states, one of the most relevant uses
of mixed-state entanglement is as an input to entanglement
distillation protocols \cite{Bennett1996,HHHH09}. Usefulness for such a
task is measured by distillable entanglement \cite{BDSW96}
$E_D(\rho)$, which quantifies how many pure Bell states can be obtained from
many copies of $\rho$ by local operations and classical communication
(per copy and in the limit of
many copies). While $E_D(\rho)>0$ for all entangled states of two
qubits, in general only upper and lower bounds are known.
We use $E_{D\to}$, the entanglement that can be distilled using only
one-way communication and which is given by \cite{DeWi03}
$E_{D\to}(\rho)=\max\left\{ 0,S(\rho_1)-S(\rho),S(\rho_2)-S(\rho)
\right\}$, where $S$ is the von Neumann entropy and $\rho_i$ the reduced
state at site $i=1,2$. 
Using this lower bound we find that the steady
states in the \textit{continuous} QHE-based protocols are too noisy to contain meaningful one-way
distillable entanglement ($E_{D\to}(\rho_s)<0.01$), while the \textit{stroboscopic} SAW-based
schemes produce $0.1-0.2$ebits of $E_{D\to}$, cf. Fig.~\ref{fig:ED},
showing that from a supply of $5n-10n$ such pairs we can distill $n$
high-fidelity Bell states which would, in turn, allow for, e.g.,
quantum teleportation or remote gate implementation.
Similar considerations should apply for stroboscopic QHE-based settings with accurate control over the electron dwell times, 
as experimentally demonstrated for example in Ref.~\citep{Bocquillon2014} .

%-----------------------------------------------------------------------------
\section{Conclusions}
\label{sec:conclusions}
To conclude, we have presented a general scheme for the deterministic generation of entanglement between spins confined in spatially separated gate-defined QDs. 
We have detailed our ideas for two specific electron-based setups feasible with current state-of-the-art technology, for which the coherence length of the corresponding quantum channels should allow us to generate sizable entanglement ($E_F\approx 0.2-0.7$) over distances of up to $100\mu$m. 
While such noisy, modestly entangled two-qubit states can be used, e.g., for quantum teleportation, their main use lies in the fact that they can be distilled into highly entangled states by means of local operations on several copies \citep{Bennett1996, Auer2015}. We have seen, in particular, that the stroboscopic schemes generate a sizable amount of distillable entanglement.
Running our steady-state scheme on several spin qubits in parallel could provide deterministic inputs to such a distillation procedure.
We have focused on
GaAs-based systems, as these have been investigated most thoroughly in experiments,
with the ambient nuclei posing one of the dominant sources of undesired noise.  
Two complementary strategies to address the role of nuclear spins 
in future studies 
would be (i) either to investigate nuclear-spin-free systems with $T_{2}^{*}{>}100\mu\mathrm{s}$
\citep{Hamaya2006, Veldhorst2014} or (ii) to associate the Heisenberg coupling (\ref{eq:heisenberg}) with the hyperfine interaction between ancilla electron spins and collective nuclear spin operators, 
with (possibly large) collective spin operators $\mathbf{I}_{i}(i=1,2)$ replacing the spin-$1/2$ system electron spins $\mathbf{S_{i}}$ considered in this work. 
By carefully choosing the spin-projection of the injected ancilla spins as well as the interaction times between electron and nuclear spins via the dwell times of the ancilla electrons in the QDs, one should be able to engineer a dissipative master equation of the form given in Eq.~(\ref{eq:goal1}), 
again with the replacement $\mathbf{S}_{i} \rightarrow \mathbf{I}_{i}$. 
Since nuclear spin ensembles typically comprise $10^{4}-10^{6}$ nuclei, 
this scheme could possibly generate large amounts of entanglement over mesoscopically large distances, 
provided that narrowed nuclear spin states with a width much smaller than the average polarization are prepared initially \citep{Schuetz2013}.

%-----------------------------------------------------------------------------
\begin{acknowledgments}
\textit{Acknowledgments.---}M.B. thanks the theory division of the Max Planck Institute of Quantum Optics for their hospitality. 
M.B. and M.J.A.S. would like to thank A. Gonzalez-Tudela for fruitful discussions.
G.G. (M.B.
and G.P.) acknowledges support by the Ministerio de Economia y
Competitividad through Project No.
FIS2014-55987-P (MAT2014-58241-P). M.J.A.S., G.G.,
and J.I.C. acknowledge support by the European
Commission via project SIQS and by the Deutsche Forschungsgemeinschaft
within the Cluster of Excellence NIM. 
\end{acknowledgments}

\appendix
%-----------------------------------------------------------------------------
\section{Entanglement of formation\label{sec:EOF}}
The entanglement measure used in this work is the entanglement
of formation ($E_F$) \citep{Wootters1998}, defined as the minimum average
entanglement of an ensemble of pure states that represents the mixed
state $\rho$. It quantifies the necessary resources to create a given
entangled state. For a mixed state $\rho$ of two qubits the concurrence
is ${\cal C}=\text{max}\left\{ 0,\lambda_{1}-\lambda_{2}-\lambda_{3}-\lambda_{4}\right\} $,
where $\lambda_{i}$ are the square roots of the eigenvalues of the
matrix $\rho A\rho^{*}A$ arranged in decreasing order, where $A$
is the antidiagonal matrix with elements $\left\{ -1,1,1,-1\right\} $.
For two qubits it ranges from 0 (separable states) to 1 (maximally entangled states).
The $E_F$ can be calculated from the concurrence
as
\begin{eqnarray}
E_F & = & -\frac{1+\sqrt{1-{\cal C}^{2}}}{2}\log_{2}\frac{1+\sqrt{1-{\cal C}^{2}}}{2}\nonumber \\
 && -  \frac{1-\sqrt{1-{\cal C}^{2}}}{2}\log_{2}\frac{1-\sqrt{1-{\cal C}^{2}}}{2}\label{eq:EOF}
\end{eqnarray}
and also ranges from $0$ to $1$.

%-----------------------------------------------------------------------------
\section{Cascaded master equation for ancilla system}\label{sec:cascaded-meq}

In Appendix~\ref{sec:input-output} we introduce the fermionic input-output formalism
\citep{Gardiner2004} and apply it to ``cascaded quantum
systems'', which consist of quantum nodes connected through an ideal chiral
reservoir. Then in Appendix~\ref{sec:ancilla} we employ the obtained cascaded master equation (ME) to model the ancilla quantum dots (QDs) connected via a quantum Hall edge (QHE) state as considered in the main text.

%-----------------------------------------------------------------------------
\subsection{Fermionic input-output formalism} \label{sec:input-output}

First of all, we address the interaction of a system with
a Markovian reservoir of non-interacting fermions. The total Hamiltonian
has the generic system Hamiltonian $H_{\text{S}}$, the bath Hamiltonian
\begin{equation}
H_{B}=\int_{0}^{\infty}d\omega\omega f^{\dagger}(\omega)f(\omega)\ ,\label{eq:bath}
\end{equation}
where $\omega$ is the bath energy and $f(\omega)$ are bath fermionic annihilation
operators with anticommutation relations $\left[f(\omega),f(\omega')^{\dagger }\right]_+=\delta(\omega-\omega')$, and the interaction Hamiltonian
\begin{equation}
H_{\text{SB}}=i\int_{0}^{\infty}d\omega\sqrt{\frac{\gamma}{2\pi}}\left\{ f^{\dagger}(\omega)d-d^{\dagger}f(\omega)\right\} \ ,\label{eq:interaction}
\end{equation}
where $d$ is a fermionic annihilation operator acting on the
system and the coupling to the reservoir
is assumed to be independent of the frequency (Markov approximation). The Heisenberg equation
of motion of the bath operators is
\begin{equation}
\dot{f}(\omega)=-i\omega f(\omega)+\sqrt{\frac{\gamma}{2\pi}}d\ ,\label{eq:f(omega)1}
\end{equation}
which can be formally integrated as
\begin{equation}
f(\omega)=e^{-i\omega t}f(\omega,0)+\sqrt{\frac{\gamma}{2\pi}}\int_{0}^{t}dt'e^{-i\omega(t-t')}d(t')\ .\label{eq:f(omega)2}
\end{equation}
Here $f\left(\omega,0\right)$ is the value of $f\left(\omega\right)$
at time $t=0$. A general system operator $a$ may commute or anticommute
with the bath operators depending on its nature. We call it if even
if it commutes with all bath operators and odd if not.
The Heisenberg equation of motion is
\begin{eqnarray}
\dot{a} & = & -\frac{i}{\hbar}\left[a,H_{\text{S}}\right]\label{eq:a1}\\
 & + & \int_{0}^{\infty}d\omega\sqrt{\frac{\gamma}{2\pi}}\left\{ \mp f^{\dagger}(\omega)\left[a,d\right]_{\pm}-\left[a,d^{\dagger}\right]_{\pm}f(\omega)\right\} \ ,\nonumber 
\end{eqnarray}
where the top (bottom) signs apply for odd (even) $a$ operator and $\left[A,B\right]_{\pm}=AB\pm BA$. Inserting
the expression~(\ref{eq:f(omega)2}) into Eq.~(\ref{eq:a1})
we derive the quantum Langevin equation 
\begin{eqnarray}
\dot{a} & = & -\frac{i}{\hbar}\left[a,H_{\text{S}}\right]\mp\left\{ \sqrt{\gamma}f_{\text{in}}^{\dagger}(t)+\frac{\gamma}{2}d^{\dagger}(t)\right\} \left[a,d\right]_{\pm}\nonumber \\
 &  & -\left[a,d^{\dagger}\right]_{\pm}\left\{ \sqrt{\gamma}f_{\text{in}}(t)+\frac{\gamma}{2}d(t)\right\} \ ,\label{eq:a2}
\end{eqnarray}
where
\begin{equation}
f_{\text{in}}(t)=\frac{1}{\sqrt{2\pi}}\int_{0}^{\infty}d\omega e^{-i\omega t}f(\omega,0)\label{eq:fin}
\end{equation}
is called noise input field and is determined by the initial state of the bath.
The noise output field, defined as the time-reversed evolution from
the final time operator $f(\omega,t_{\text{f}})$, is related to it
by
\begin{equation}
f_{\text{out}}(t)-f_{\text{in}}(t)=\sqrt{\gamma}d(t)\ ,\label{eq:out-in}
\end{equation}
an identity known as the input-output relation. Up to this point, no
assumption has been made concerning the density operator of the bath.
We will use the  white-noise approximation which assumes the following correlation
functions for the input field: $\left\langle f_{\text{in}}^{\dagger}\left(\omega\right)f_{\text{in}}\left(\omega'\right)\right\rangle =\bar{N}\delta\left(\omega-\omega'\right)$
and $\left\langle f_{\text{in}}\left(\omega\right)f_{\text{in}}^{\dagger}\left(\omega'\right)\right\rangle =\left(1-\bar{N}\right)\delta\left(\omega-\omega'\right)$.
Here $\bar{N}$ is the Fermi distribution function of a thermal reservoir. 
Moreover we will assume a weak system-reservoir coupling in the
sense that the correlation functions of the bath are not affected
by the interaction.

The input-output formalism provides a powerful treatment for two or
more subsystems sharing a common unidirectional reservoir \citep{H.J.Carmichael1993,Gardiner1993,Kolobov87},
also known as cascaded quantum systems. Let us consider the case of
two nodes coupled to the reservoir via Eq.~(\ref{eq:interaction})
with operators $d_{j}(j=1,2)$. Following the previous argument a
system operator of subsystem $j$, $a_{j}$, follows the Eq.~(\ref{eq:a2}) with the change $d\rightarrow d_{j}$, $\gamma\rightarrow\gamma_{j}$
and $f_{\text{in}}\rightarrow f_{\text{in}}^{(j)}$. The fact that
the reservoir is common and unidirectional implies a relation between
the output of subsystem 1 and the input in 2. For a dispersion-free
channel $f_{\text{in}}^{(2)}(t)=f_{\text{out}}^{(1)}(t-L/v)$, where $L$ is the
distance between the two subsystems and $v$ the group velocity of
the reservoir modes, i.e., all the output of the first subsystem is used
later as the input into the second one, therefore we are able to write
a generic equation for an odd (even) operator as \citep{Stace2004}\begin{widetext}
\begin{eqnarray}
\dot{a}(t) & = & -\frac{i}{\hbar}\left[a,H_{\text{S}}\right]\mp\left\{ \sqrt{\gamma_{1}}f_{\text{in}}^{\dagger(1)}(t)+\frac{\gamma_{1}}{2}d_{1}^{\dagger}(t)\right\} \left[a,d_{1}\right]_{\pm}-\left[a,d_{1}^{\dagger}\right]_{\pm}\left\{ \sqrt{\gamma_{1}}f_{\text{in}}^{(1)}(t)+\frac{\gamma_{1}}{2}d_{1}(t)\right\} \nonumber \\
 & \mp & \left\{ \sqrt{\gamma_{2}}f_{\text{in}}^{(1)\dagger}(t-L/v)+\frac{\gamma_{2}}{2}d_{2}^{\dagger}(t)+\sqrt{\gamma_{1}\gamma_{2}}d_{1}^{\dagger}(t-L/v)\right\} \left[a,d_{2}\right]_{\pm}\nonumber \\
 & - & \left[a,d_{2}^{\dagger}\right]_{\pm}\left\{ \sqrt{\gamma_{2}}f_{\text{in}}^{(1)}(t-L/v)+\frac{\gamma_{2}}{2}d_{2}(t)+\sqrt{\gamma_{1}\gamma_{2}}d_{1}(t-L/v)\right\} \ .\label{eq:a3}
\end{eqnarray}
\end{widetext}Since the coupling operators $d_{1,2}$ are fermionic annihilation (odd) operators, they (anti)commute with any (odd) even operator $a$ of the other system. Then it is clear from Eq.~(\ref{eq:a3}) that the time evolution of an operator
of the second subsystem depends on the first one but not the other
way around, which reflects the unidirectionality condition. Following \citep{Gardiner1993,Cirac1997}, for a dispersionless channel, the fixed time delay may be set to zero, i.e., one can choose $L/v=0^{+}$ without loss of generality.
The previous equation can be easily
rewritten as
\begin{eqnarray}
\dot{a}(t) & = & -\frac{i}{\hbar}\left[a,H_{\text{S}}+\frac{i\sqrt{\gamma_{1}\gamma_{2}}}{2}\left(d_{1}^{\dagger}d_{2}-d_{2}^{\dagger}d_{1}\right)\right]\label{eq:a4}\\
 & - & \left[a,d^{\dagger}\right]_{\pm}\left\{ \frac{d}{2}+f_{\text{in}}^{(1)}(t)\right\} \mp\left\{ \frac{d^{\dagger}}{2}+f_{\text{in}}^{\dagger(1)}(t)\right\} \left[a,d\right]_{\pm}\nonumber 
\end{eqnarray}
in terms of the nonlocal operator $d=\sqrt{\gamma_{1}}d_{1}+\sqrt{\gamma_{2}}d_{2}$.
Once we have derived this quantum Langevin equation, we can find a
ME for the partial density operator excluding the bath
$\varrho$ by tracing out the bath degrees of freedom from the total
density operator ${\cal W}$, $\varrho=\text{tr}_{\text{B}}\left\{ {\cal W}\right\} $.
For this we make use of the relation $\text{tr}\left\{ \dot{a}(t){\cal W}\right\} =\text{tr}\left\{ a\dot{{\cal W}}(t)\right\} =\text{tr}_{\text{s}}\left\{ a\dot{\varrho}(t)\right\} $.
Since any physical state is fully described by the expectation values
of even observables (the odd ones have vanishing expectation value
due to the parity superselection rule) we can restrict ourselves in Eq.~(\ref{eq:a4})
to the lower sign for all observables of interest and end up with
the ME 
\begin{eqnarray}
\dot{\varrho} & = & -i\left[H_{\text{S}}+\frac{i\sqrt{\gamma_{1}\gamma_{2}}}{2}\left(d_{1}^{\dagger}d_{2}-d_{2}^{\dagger}d_{1}\right),\varrho\right]\nonumber \\
 & + & \frac{1}{2}\left(1-\bar{N}\right){\cal D}\left[d\right]\varrho+\frac{1}{2}\bar{N}{\cal D}\left[d^{\dagger}\right]\varrho\ ,\label{eq:cascmeq}
\end{eqnarray}
where ${\cal D}[A]\varrho=2A\varrho A^{\dagger}-A^{\dagger}A\varrho-\varrho A^{\dagger}A$
and $\bar{N}$ is the Fermi distribution function of the fermionic
reservoir. This expression contains the nonlocal coherent and incoherent
contributions of the coupling between subsystems mediated by the reservoir.
For simplicity we have neglected the spin index in this derivation.
Moreover,
in the main text we work in a rotating frame such that the global
homogeneous magnetic field drops out. If the ancilla dots energy levels
are not aligned within $\gamma$, this would generate an undesired
rotation of the nonlocal terms in Eq.~(\ref{eq:cascmeq}) \citep{Stace2004}.

%-----------------------------------------------------------------------------
\subsection{Ancilla quasisteady state} \label{sec:ancilla}

The dynamics of the ancilla QDs connected via a QHE state considered in the main text can be described by Eq.~(\ref{eq:cascmeq}). Note that we consider only the nearest resonant
subband because the tunneling rates decrease exponentially with the distance
from the dots \citep{Stace2004}. For simplicity,  we restrict ourselves to the case $\gamma\equiv\gamma_1=\gamma_2$. Moreover, we consider the case of an empty channel $\bar{N}=0$, we need to account explicitly for spins  and we add the contribution from the reservoir that pumps electrons into the first ancilla QD. Finally, if the  spin-resolved  levels of the two ancilla QDs are aligned, the system Hamiltonian term vanishes in a suitable rotating frame. Therefore the dynamics of the ancilla dots is described by the transport Liouville superoperator ${\cal L}_{\text{tr}}\varrho=\sum_{\sigma} {\cal L}_{\text{tr},\sigma}\varrho$ with
\begin{eqnarray}
{\cal L}_{\text{tr},\sigma}\varrho & = & \frac{\gamma_{\text{L},\sigma}}{2}{\cal D}\left[d_{1\sigma}^{\dagger}\right]\varrho+\frac{\gamma}{2}{\cal D}\left[d_{1\sigma}+d_{2\sigma}\right]\varrho\nonumber \\
 & + & \frac{\gamma}{2}\left[d_{1\sigma}^{\dagger}d_{2\sigma}-d_{2\sigma}^{\dagger}d_{1\sigma},\varrho\right]. \label{eq:cascaded-ap}
\end{eqnarray}

\begin{figure}
\includegraphics[width=1\columnwidth]{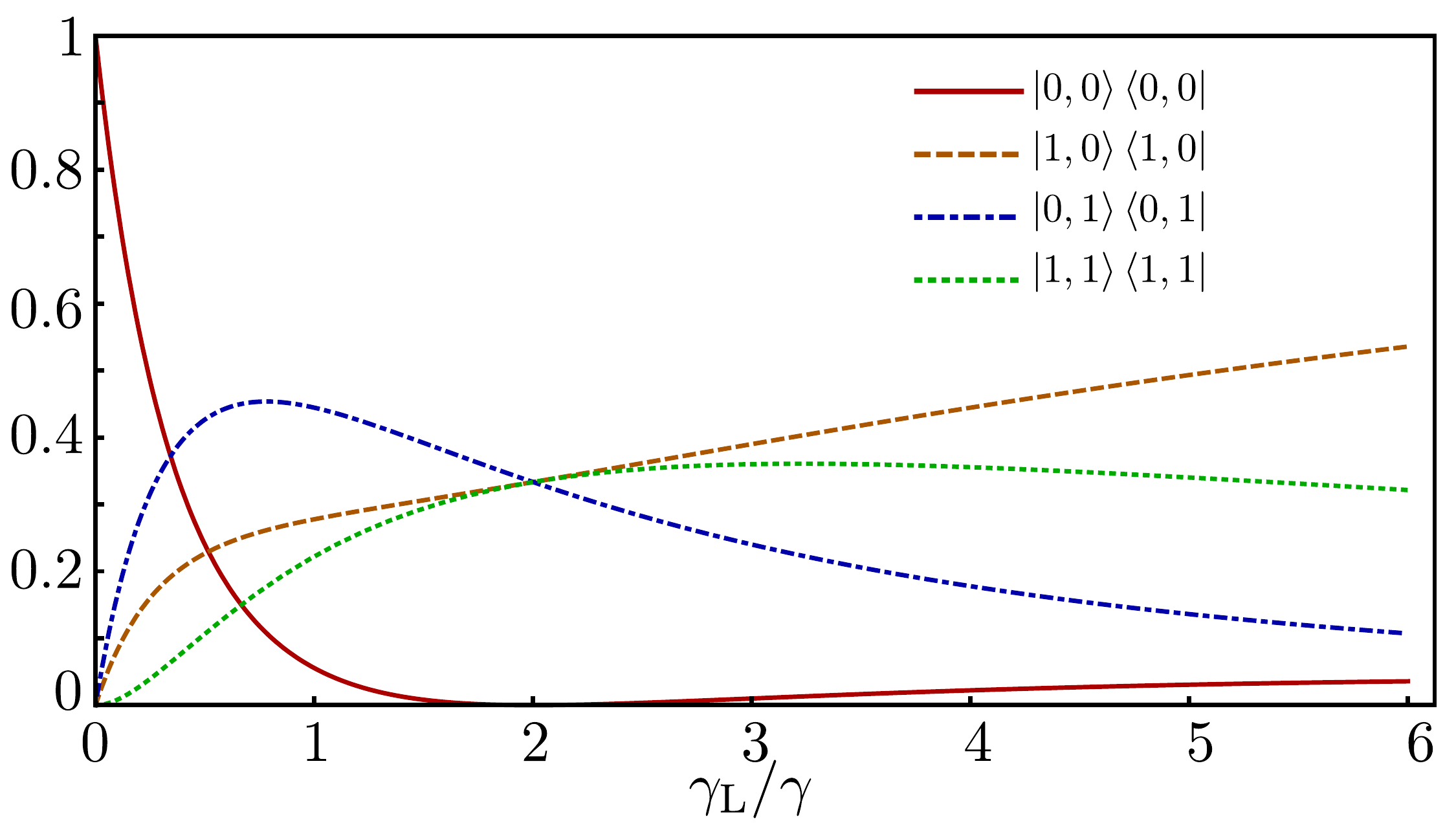}

\protect\caption{\label{fig:populations}Value of the diagonal elements of the ancilla  steady state in Eq.~(\ref{eq:steady-state}) as a function of the ratio $\gamma_{\text{L}}/\gamma$.
}
\end{figure}

For fast dissipation ($\gamma,\gamma_{\text{L}}\gg J$), the auxiliary dots settle into a quasisteady state ($\rho_{\text{a}}^{\text{ss}}$) on a time scale much shorter
than the relevant system dots dynamics. We now compute and
analyze this quasisteady state since it will play a central role for the
system dots ME to be derived in Appendix~\ref{sec:Effective-master-equation}.  
If a single spin component is introduced, $\gamma_{\text{L},\downarrow}=0$ and $\gamma_{\text{L}}\equiv\gamma_{\text{L},\uparrow}$, the quasisteady state associated with Eq.~(\ref{eq:cascaded-ap}) is
\begin{widetext}
\begin{eqnarray}
\rho_{\text{a}}^{\text{ss}} & = & \frac{1}{\left(\gamma_{\text{L}}+\gamma\right)\left(\gamma_{\text{L}}+2\gamma\right)^2} \left\{\gamma\left(2\gamma-\gamma_{\text{L}}\right)^2\left|0,0\right>\left<0,0\right|+\gamma_{\text{L}}\left(4\gamma^2+\gamma_{\text{L}}^2\right)\left|\uparrow,0\right>\left<\uparrow,0\right|+8\gamma_{\text{L}}\gamma^2\left|0,\uparrow\right>\left<0,\uparrow\right|\right.\nonumber \\
 & - &\left.2\gamma\gamma_{\text{L}}\left(\gamma_{\text{L}}+2\gamma\right)\left( \left|\uparrow,0\right>\left<0,\uparrow\right|+\left|0,\uparrow\right>\left<\uparrow,0\right| \right)+4\gamma\gamma_{\text{L}}^2\left|\uparrow,\uparrow\right>\left<\uparrow,\uparrow\right|\right\} . \label{eq:steady-state}
\end{eqnarray}
\end{widetext}The average populations of the ancilla dots depend on the reservoir and channel rates  as shown in Fig.~\ref{fig:populations}.

For all
$\gamma,\gamma_L\not=0$, the quasisteady state is entangled (due to the Markovian
coupling to the common channel) and reaches an $E_F$ of $\sim 0.55$ at
$\gamma_L=2\gamma$, at which point the steady state is a mixture of
the two-electron state $\left| \uparrow,\uparrow\right>$ and the maximally entangled
state $\left|\uparrow,0\right>-\left|0,\uparrow\right>$ that is a ``dark state'' for the
collective coupling  via the operator $(d_{1,\uparrow}+d_{2,\uparrow})$ in
Eq.~(\ref{eq:cascaded-ap}).  
However, this entanglement comes in a form of limited usefulness as it involves a superposition of a single fermion in
the first or in the second ancilla and due to fermionic superselection rules a
single such state (while entangled
\citep{Banuls2007,2015arXiv151104450D}) cannot be distinguished from a
separable state by local operations. 
Our scheme shows that this entanglement can still provide the quantum correlations necessary to produce a usable spin-qubit entanglement for the system spins, which are weakly coupled to this ancilla system.

In accordance with the cascaded nature of the system, $\varrho$ in Eq.~(\ref{eq:cascaded-ap})
takes into account a time delay between systems 1 and 2. If transport happens almost instantaneously even on the time-scale of the channel-ancilla coupling ($L/v\ll 1/\gamma$), the delay can be neglected and the quasisteady state in Eq.~(\ref{eq:steady-state}) can be understood as an equal-times state. However, this condition limits the length of the edge channels to $L<1\mu$m. 
For larger separations ($L/v\gg 1/\gamma$) we see that the first QD is
driven into its steady state before the electrons that interact with it
have time to reach the second QD. Hence we conclude that at any given time,
QD1 and QD2 are not entangled; instead, QD1 is getting entangled with
the bath (the electron modes in the channel connecting the two QDs).
This notwithstanding, as the cascaded equation tells us, this
system-bath entanglement is faithfully transported to QDs so that
time-delayed measurements at the two dots show strong quantum
correlations. If other quantum systems (such as the system spins in our setup)
interact weakly with these two correlated ancillas they are exposed to
an nonlocal master equation that can be
effectively taken as an equal-time equation if $L/v$ is short compared
to the time scale of the qubit dynamics, shown in Appendix~\ref{sec:Adiabatic-Elimination-and-effective-equation}  to be on
the order of $J_0^2/\gamma$. For realistic parameter values, we thus
obtain a standard equal-time entangled steady state for channel lengths
of up to a few tens of micrometers.

%-----------------------------------------------------------------------------
\section{Adiabatic Elimination of the ancilla system \label{sec:Adiabatic-Elimination-and-effective-equation}}

%-----------------------------------------------------------------------------
\subsection{Adiabatic Elimination \label{sec:Adiabatic-Elimination}}

The adiabatic elimination is a useful method when one has a main system
weakly coupled to an auxiliary system, which undergoes fast dynamics
(given by a Liouvillian ${\cal L}_{0}$), since it allows us to determine the
effective dynamics of the main system to (in principle) arbitrary order in the interaction \citep{Kessler2012}.
Analogously to the Schrieffer-Wolff transformation for closed systems,
it allows us to decouple the slow subspace, given by the steady state
of the auxiliary system, i.e., ${\cal L}_{0}\rho_{\text{a}}^{\text{ss}}=0$ \cite{Note3}%\footnote{We assume that ${\cal{L}}_0$ has a unique steady state $\rho_{\text{a}}^{\text{ss}}$}
,
from the fast one. To this end, one defines the projector ${\cal P}$
by its action over the total density matrix (DM) ${\cal P}\varrho=\text{tr}_{\text{a}}\left\{ \varrho\right\} \otimes\rho_{\text{a}}^{\text{ss}}=\rho\otimes\rho_{\text{a}}^{\text{ss}}$,
where we have introduced the reduced DM as the trace over
the auxiliary system $\rho\equiv\text{tr}_{\text{a}}\left\{ \varrho\right\} $,
and apply it to the total ME of the form $\dot{\varrho}=\left({\cal L}_{0}+{\cal V}\right)\varrho$,
where ${\cal V}$ is the perturbative part. In this way we can obtain the subsequent
orders of the effective Liouville operator expansion that governs
the dynamics of the main system ($\dot{\rho}=\text{tr}_{\text{a}}\left\{ L_{\text{eff}}\varrho\right\} $)
\citep{Kessler2012}. Defining the Laplace transform of ${\cal L}_{0}$
as ${\cal L}_{0}^{-1}=-\int_{0}^{\infty}d\tau e^{{\cal L}_{0}\tau}$,
one can easily find
\begin{eqnarray}
L_{\text{eff,1}} & = & {\cal P}{\cal V}{\cal P}\ ;\label{eq:Leff1}\\
L_{\text{eff,2}} & = & -{\cal P}{\cal V}{\cal Q}{\cal L}_{0}^{-1}{\cal Q}{\cal V}{\cal P}\ ;\label{eq:Leff2}
\end{eqnarray}
where ${\cal Q}=1-{\cal P}$ is the projector into the fast subspace.
The perturbation ${\cal V}$ contains the interaction between the
main and auxiliary systems as well as a main-system Hamiltonian, i.e., in general
\begin{equation}
{\cal V}\varrho=-i\sum_{j=1}^{N}\left[A_{j}\otimes S_{j},\varrho\right]-i\sum_{j=1}^{N}a_{j}\left[S_{j},\varrho\right]\ .\label{eq:perturbation}
\end{equation}
Here $A_{j}$ and $S_{j}$ are auxiliary and main-system operators, respectively,
and $a_{j}\in {\rm I\!R}$. The first-order term of $\dot{\rho}$ is
\begin{equation}
\text{tr}_{\text{a}}\left\{ L_{\text{eff,1}}\varrho\right\} =-i\sum_{j=1}^{N}\left[\left\langle A_{j}\right\rangle _{\text{ss}}S_{j},\rho\right]-i\sum_{j=1}^{N}a_{j}\left[S_{j},\rho\right]\ ,\label{eq:firstorder}
\end{equation}
which means that to first order the main system experiences the effect of
the mean values of the auxiliary-system operators in the quasisteady
state, $\left\langle A_{j}\right\rangle _{\text{ss}}=\text{tr}_{\text{a}}\left\{ A_{j}\rho_{\text{a}}^{\text{ss}}\right\} $,
plus the original main-system Hamiltonian. To second order, one can
show
\begin{eqnarray}
\text{tr}_{\text{a}}\left\{ L_{\text{eff,2}}\varrho\right\}  & = & -\sum_{i,j}\text{tr}_{\text{a}}\left\{ \delta A_{i}{\cal L}_{0}^{-1}\delta A_{j}\rho_{\text{a}}^{\text{ss}}\right\} \left[S_{j}\rho,S_{i}\right]\nonumber \\
 & - & \sum_{i,j}\text{tr}_{\text{a}}\left\{ \delta A_{i}{\cal L}_{0}^{-1}\rho_{\text{a}}^{\text{ss}}\delta A_{j}\right\} \left[S_{i},\rho S_{j}\right]\ ,\label{eq:trLeff2}
\end{eqnarray}
where $\delta A_{j}$ are the fluctuations of the auxiliary-system
operators: $\delta A_{j}=A_{j}-\left\langle A_{j}\right\rangle _{\text{ss}}$.
Using the quantum regression theorem
\begin{eqnarray}
\text{tr}_{\text{\text{a}}}\left\{ \delta A_{i}e^{{\cal L}_{0}\tau}\left[\delta A_{j}\rho_{\text{a}}^{ss}\right]\right\}  & = & \left\langle \delta A_{i}(\tau)\delta A_{j}\right\rangle _{\text{ss}}\ ;\nonumber \\
\text{tr}_{\text{a}}\left\{ \delta A_{i}e^{{\cal L}_{0}\tau}\left[\rho_{\text{a}}^{ss}\delta A_{j}\right]\right\}  & = & \left\langle \delta A_{j}\delta A_{i}(\tau)\right\rangle _{\text{ss}}\ ;\label{eq:QRT}
\end{eqnarray}
and  the relation
 $\left\langle \delta A_{j}\delta A_{i}(\tau)\right\rangle _{\text{ss}}^{*}=\left\langle \delta A_{i}^{\dagger}(\tau)\delta A_{j}^{\dagger}\right\rangle _{\text{ss}}$,
Eq.~(\ref{eq:Leff2}) reads
\begin{eqnarray}
\text{tr}_{\text{a}}\left\{ L_{\text{eff,2}}\varrho\right\}  & = & \sum_{i,j}{\cal C}\left(A_{i},A_{j}\right)\left[S_{j}\rho,S_{i}\right] \nonumber \\
 & + & \sum_{i,j} {\cal C}^*\left(A_{i}^{\dagger},A_{j}^{\dagger}\right)\left[S_{j}^{\dagger}\rho,S_{i}^{\dagger}\right]^{\dagger}\ ,\label{eq:trLeff2-2}
\end{eqnarray}
where we introduce the correlation functions
\begin{eqnarray}
{\cal C}\left(A_{i},A_{j}\right) & =&\text{tr}_{\text{a}}\left\{ \delta A_{i}{\cal L}_{0}^{-1}\delta A_{j}\rho_{\text{a}}^{\text{ss}}\right\} \ .\label{eq:Corr}
\end{eqnarray}
In the specific case under consideration in the main text, ${\cal L}_{0}={\cal L}_{\text{tr}}$
and ${\cal V}\varrho=-i\left[H_{\text{Z}}+H_{\text{IN}},\varrho\right]$. 
\\

%-----------------------------------------------------------------------------
\subsection{\label{sec:Effective-master-equation}Effective master equation for
the system spins}

In the following, we apply the method of adiabatic elimination developed
in Appendix~\ref{sec:Adiabatic-Elimination} to the physical setup based on QHE states in order to eliminate the ancilla coordinates
and obtain an effective ME for the system spins. 
An electron occupying the ancilla dot $j$ interacts locally with the system spin $\mathbf{S}_{i}$ via the Heisenberg exchange interaction \citep{Loss1998}
\begin{eqnarray}
H_{\text{IN}}^{i,j} & = & J_{i,j} \mathbf{S}_{i} \cdot \boldsymbol{\sigma}_{j}, \label{eq:heisenberg-ap}
\end{eqnarray}
where $\boldsymbol{\sigma}_{j} = \frac{1}{2} \sum_{\sigma,\sigma'}d_{j\sigma}^{\dagger} \boldsymbol{\tau} d_{j\sigma}$ refers to the spin-$1/2$ ancilla operator; here, $d_{j\sigma}^{\dagger}$ creates an electron with spin $\sigma=\uparrow,\downarrow$ in the ancilla dot $j$ and $ \boldsymbol{\tau}$ is the vector of Pauli matrices.
The complete interaction Hamiltonian is then $H_{\text{IN}}=\sum_{\left<i,j\right>} H_{\text{IN}}^{i,j}$, which describes local spin-spin interactions between ancilla and system dots. 
According to Eq.~(\ref{eq:perturbation}), the generic ancilla operators $A_n$ are $\sigma_j^{\alpha}$, with $\alpha=x,y,z$ and $j=1,...4$,  and the system operators $S_n$ are $J_{i,j}S_i^{\alpha}$, with $i=1,2$.

\begin{figure}
\includegraphics[width=1\columnwidth]{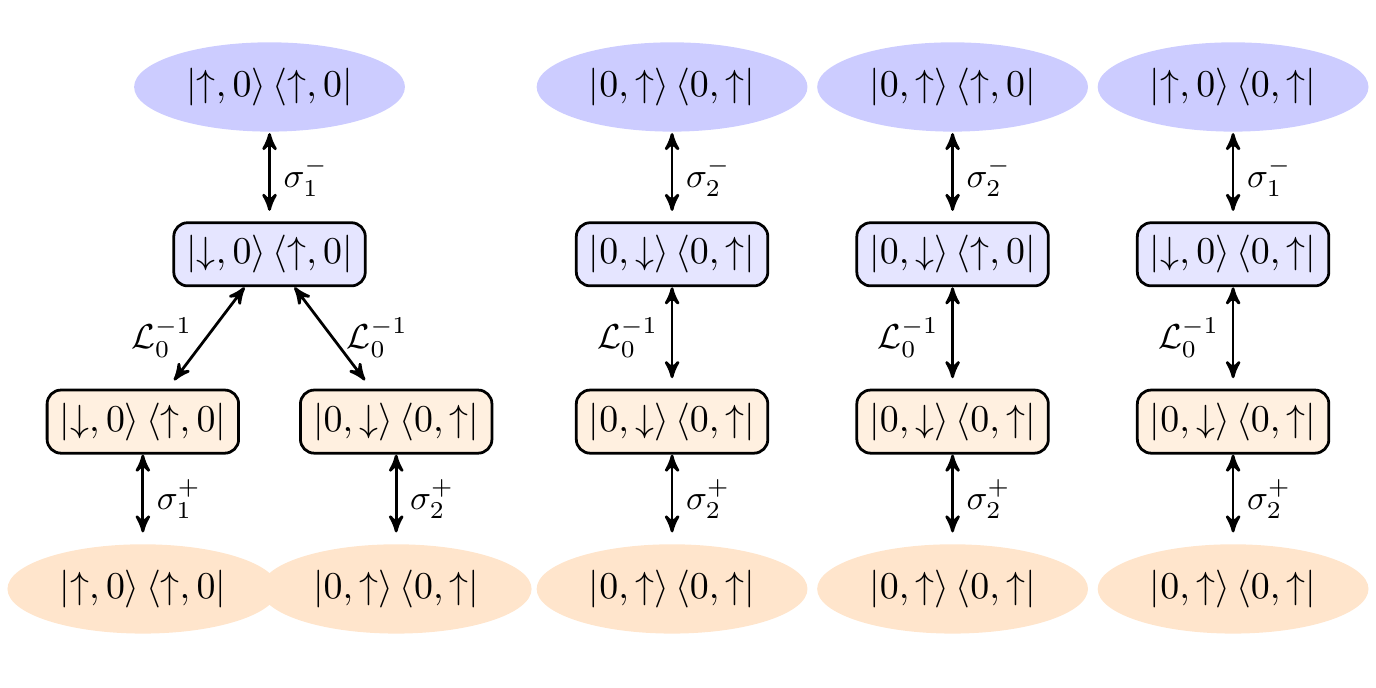}

\protect\caption{\label{fig:Ad-elim}Schematic representation of the second-order correlation
functions; compare Eq.~(\ref{eq:Corr}). The different components of $\rho_{\text{a}}^{\text{ss}}$ are coupled to the elements in rectangles via ancilla spin-flip operators $\sigma_{1,2}^{-}$.  Then, the pseudoinverse of the transport Liouvillian ${\cal L}_{\text{tr}}$
couples them to the matrix elements shown in the bottom rectangles. Finally, a second application of the ancilla spin-flip operators couples the initial component to the components shown in the bottom ellipses. 
For simplicity, this example
refers to the limiting case $\gamma_{\text{L}}\ll\gamma$,;
%chosen for simplicity
%\textcolor{blue}{state with one electron in each auxiliary QD is not populated.}
%two-electrons state population is zero.
in this regime one can restrict the discussion to the single-electron regime, where at most a single electron is found in the ancilla system (comprising the two ancilla dots) and the population of the state with one electron in each of the two auxiliary QD is negligibly small; moreover, double occupation of a single ancilla QD is disregarded due to strong Coulomb interaction effects. Note that this schematic representation refers to just two system QDs coupled to just two ancilla dots interconnected by a single channel.
}
\end{figure}

According
to Eq.~(\ref{eq:firstorder}) the first-order contributions are given
by the mean value of the magnetic field created by the ancilla electrons
in the quasisteady state $\rho_{\text{a}}^{\text{ss}}$, i.e., $\left\langle \sigma_{i}^{z}\right\rangle _{\text{ss}}=\text{tr}_{\text{a}}\left\{ \sigma_{i}^{z}\rho_{\text{a}}^{\text{ss}}\right\} $,
and the system Hamiltonian 
\begin{eqnarray}
\text{tr}_{\text{a}}\left\{ L_{\text{eff,1}}\varrho\right\}  & = & -i\left[H_{\text{Z}}+\sum_{\left<i,j\right>}\left\langle \sigma_{j}^{z}\right\rangle _{\text{ss}}J_{i,j}S_{i}^{z},\rho\right]\ ;\label{eq:first-order-cancelation}\\
H_{\text{Z}} & = & \sum_{i}\delta_{i}S_{i}^{z}\ .\label{eq:Zeeman}
\end{eqnarray}
The local constant fields in $H_\text{Z}$  can then be chosen such that they cancel
Eq.~(\ref{eq:first-order-cancelation}) and will be on the order of the exchange coupling. 
Using Eq.~(\ref{eq:trLeff2-2}) we calculate  the second-order contribution of the coupling to two ancilla dots connected via a unidirectional channel ($J_{i}\equiv J_{i,i}$). There is a term due to the parallel component of the Heisenberg interaction ($z-z$)
\begin{eqnarray}
{\cal L}_{\text{zz}}\rho & = & \sum_{i=1}^{2}J_{i}^{2}{\cal C}\left(\sigma_{i}^{z},\sigma_{i}^{z}\right){\cal D}\left[S_{i}^{z}\right]\rho\label{eq:Lzz1}\\
 & + & J_{1}J_{2}\left({\cal C}\left(\sigma_{2}^{z},\sigma_{1}^{z}\right)+{\cal C}\left(\sigma_{1}^{z},\sigma_{2}^{z}\right)\right)\left(\left[S_{1}^{z}\rho,S_{2}^{z}\right]+\left[S_{2}^{z},\rho S_{1}^{z}\right]\right)\nonumber 
\end{eqnarray}
and another one due to the perpendicular component ($flip-flop$)
\begin{eqnarray}
{\cal L}_{\text{ff}}\rho & = & \sum_{i=1}^{2}{\cal C}\left(\sigma_{i}^{+},\sigma_{i}^{-}\right)\frac{J_{i}^{2}}{4}{\cal D}[S_{i}^{+}]\rho\label{eq:Lff1}\\
 & + & {\cal C}\left(\sigma_{2}^{+},\sigma_{1}^{-}\right)\frac{J_{1}J_{2}}{4}\left\{ \left[S_{1}^{+}\rho,S_{2}^{-}\right]+\left[S_{2}^{+},\rho S_{1}^{-}\right]\right\} .\nonumber 
\end{eqnarray}
The correlation functions are defined in Eq.~(\ref{eq:Corr}).
In Fig.~\ref{fig:Ad-elim}, we represent schematically the second-order
processes related to the operators $\sigma_{i}^{\pm}$. Note that the unidirectionality of the channel implies ${\cal C}\left(\sigma_{1}^{+},\sigma_{2}^{-}\right)=0$. 

%\begin{widetext}

%\end{widetext}

For practical reasons, it is more adequate to express Eqs.~(\ref{eq:Lzz1}) and~(\ref{eq:Lff1}) by means of nonlocal terms. By simply diagonalizing the
quadratic form we end up with\begin{widetext}
\begin{equation}
{\cal L}_{\text{zz}}\rho  =  \Gamma_{+}^{\text{zz}}{\cal D}[\cos\frac{\theta_{\text{zz}}}{2}S_{1}^{z}+\sin\frac{\theta_{\text{zz}}}{2}S_{2}^{z}]\rho
  +  \Gamma_{-}^{\text{zz}}{\cal D}[\sin\frac{\theta_{\text{zz}}}{2}S_{1}^{z}-\cos\frac{\theta_{\text{zz}}}{2}S_{2}^{z}]\rho\ \label{eq:Lzz}
\end{equation}
and 
\begin{equation}
{\cal L}_{\text{ff}}\rho  =  \Gamma_{+}^{\text{ff}}{\cal D}[\cos\frac{\theta_{\text{ff}}}{2}S_{1}^{+}+\sin\frac{\theta_{\text{ff}}}{2}S_{2}^{+}]\rho
  +  \Gamma_{-}^{\text{ff}}{\cal D}[\sin\frac{\theta_{\text{ff}}}{2}S_{1}^{+}-\cos\frac{\theta_{\text{ff}}}{2}S_{2}^{+}]\rho
  -  \Delta\left[S_{2}^{-}S_{1}^{+}-S_{1}^{-}S_{2}^{+},\rho\right]\ .\label{eq:ff}
\end{equation}
The rates in Eqs.~(\ref{eq:Lzz}) and ~(\ref{eq:ff}) are all
given in terms of the correlation functions as
\begin{eqnarray}
\Gamma_{\pm}^{\text{zz}} & = & \frac{1}{2}\sum_{i=1}^{2}{\cal C}\left(\sigma_{i}^{z},\sigma_{i}^{z}\right)J_{i}^{2}\pm\frac{1}{2}\sqrt{\left[{\cal C}\left(\sigma_{1}^{z},\sigma_{1}^{z}\right)J_{1}^{2}-{\cal C}\left(\sigma_{2}^{z},\sigma_{2}^{z}\right)J_{2}^{2}\right]^{2}+\left[{\cal C}\left(\sigma_{1}^{z},\sigma_{2}^{z}\right)+{\cal C}\left(\sigma_{2}^{z},\sigma_{1}^{z}\right)\right]^{2}J_{1}^{2}J_{2}^{2}};\label{eq:ratezz}\\
\Gamma_{\pm}^{\text{ff}} & = & \frac{1}{8}\sum_{i=1}^{2}{\cal C}\left(\sigma_{i}^{+},\sigma_{i}^{-}\right)J_{i}^{2}\pm\frac{1}{8}\sqrt{\left[{\cal C}\left(\sigma_{1}^{+},\sigma_{1}^{-}\right)J_{1}^{2}-{\cal C}\left(\sigma_{2}^{+},\sigma_{2}^{-}\right)J_{2}^{2}\right]^{2}+{\cal C}\left(\sigma_{2}^{+},\sigma_{1}^{-}\right)^{2}J_{1}^{2}J_{2}^{2}};\label{eq:rateff}
\end{eqnarray}
\end{widetext} and the angles that define the nonlocal operators
into the Lindblad dissipators as
\begin{eqnarray}
\theta_{\text{zz}} & = & \arctan\frac{\left({\cal C}\left(\sigma_{1}^{z},\sigma_{2}^{z}\right)+{\cal C}\left(\sigma_{2}^{z},\sigma_{1}^{z}\right)\right)J_{1}J_{2}}{{\cal C}\left(\sigma_{1}^{z},\sigma_{1}^{z}\right)J_{1}^{2}-{\cal C}\left(\sigma_{2}^{z},\sigma_{2}^{z}\right)J_{2}^{2}}\ ;\label{eq:thetaz}\\
\theta_{\text{ff}} & = & \arctan\frac{{\cal C}\left(\sigma_{2}^{+},\sigma_{1}^{-}\right)J_{1}J_{2}}{{\cal C}\left(\sigma_{1}^{+},\sigma_{1}^{-}\right)J_{1}^{2}-{\cal C}\left(\sigma_{2}^{+},\sigma_{2}^{-}\right)J_{2}^{2}}\ .\label{eq:thetaff}
\end{eqnarray}
Finally, the Hamiltonian term in Eq.~(\ref{eq:ff}) is an effective coherent
spin interaction between the spatially separated spins mediated by
the reservoir with strength
\begin{equation}
\Delta=\frac{{\cal C}\left(\sigma_{2}^{+},\sigma_{1}^{-}\right)J_{1}J_{2}}{8}\ .\label{eq:Delta}
\end{equation}

\begin{figure}
\includegraphics[width=1\columnwidth]{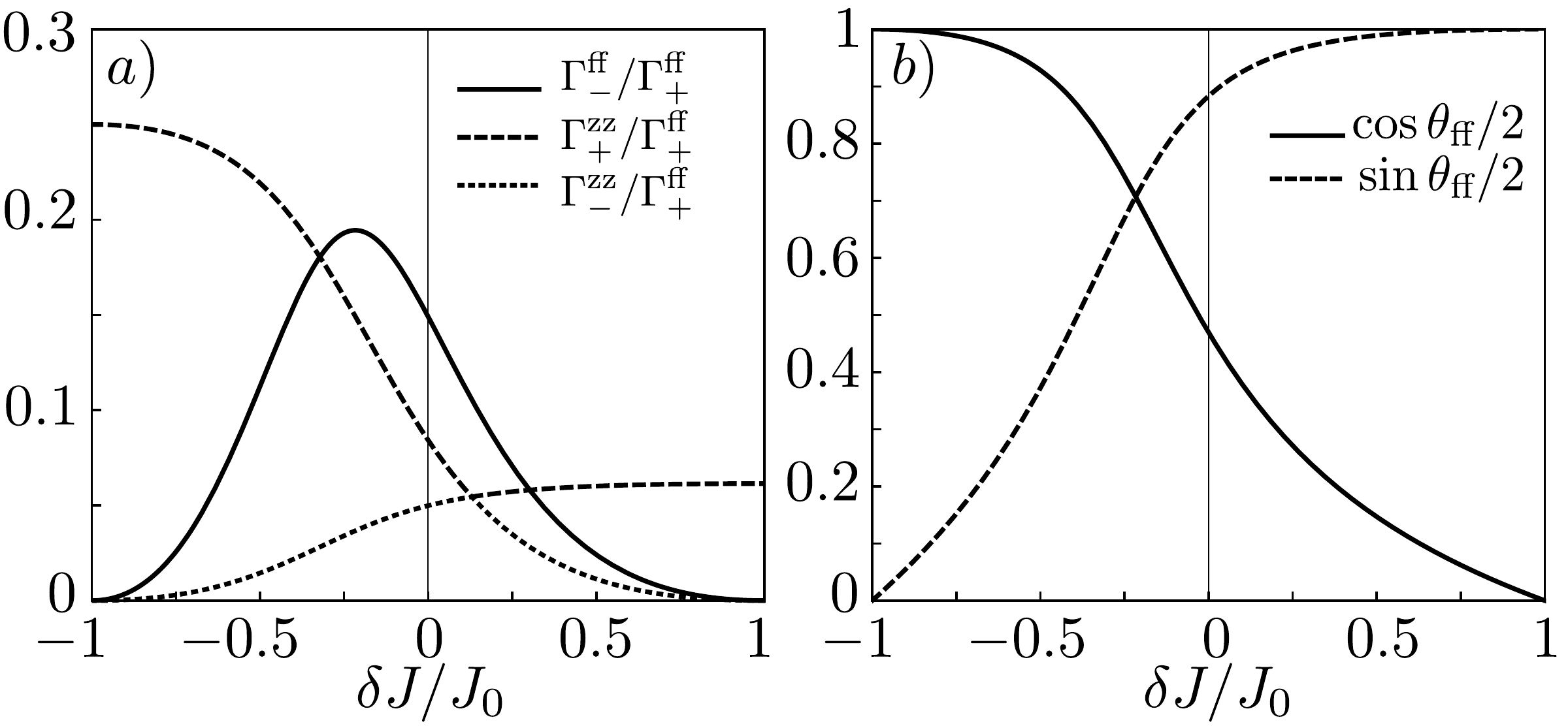}

\protect\caption{\label{fig:rates}(a) Rates of the effective ME for the
system spins. Since the rate $\Gamma_{+}^{\text{ff}}$ dominates,
we show in (b) the structure of the corresponding nonlocal operator
$\cos\frac{\theta_{\text{ff}}}{2}S_{1}^{+}+\sin\frac{\theta_{\text{ff}}}{2}S_{2}^{+}$
as a function of $\delta J$, with $J_{1(2)}=J_{0}\mp\delta J$. }
\end{figure}

Following the intuition of spin-flip processes between the localized spins and the ancilla electrons, we expect that a nonlocal
term may dominate over all other processes.   
In Fig.~\ref{fig:rates} a) 
%we show the value of the different rates contributing to Eqs.~(\ref{eq:ratezz}) and~(\ref{eq:rateff})
the different rates contributing to Eqs.~(\ref{eq:ratezz}) and~(\ref{eq:rateff}) are shown
as a function of the coupling strength difference $\delta J$,
with $J_{1(2)}=J_{0}\mp\delta J$. Clearly, the rate $\Gamma_{+}^{\text{ff}}$ is found to dominate; however, other processes may not be neglected completely.
Note that we have chosen the case of equal rates 
$\gamma_{\text{L}}=\gamma$ for simplicity because it is close to the optimum working point. For this particular case
\begin{eqnarray}
\rho_{\text{a}}^{\text{ss}} & = & \frac{1}{18} \left\{
\left|0,0\right>\left<0,0\right|+5\left|\uparrow,0\right>\left<\uparrow,0\right|+8\left|0,\uparrow\right>\left<0,\uparrow\right|\right.\nonumber \\
 & - &\left. 6\left( \left|\uparrow,0\right>\left<0,\uparrow\right|+\left|0,\uparrow\right>\left<\uparrow,0\right| \right)+4\left|\uparrow,\uparrow\right>\left<\uparrow,\uparrow\right|
 \right\} . \label{eq:steady-state-2}
\end{eqnarray}
Then the average fields are $\left\langle \sigma_{1}^{z}\right\rangle _{\text{ss}}=1/4$
and $\left\langle \sigma_{2}^{z}\right\rangle _{\text{ss}}=1/3$ and
the correlation functions are ${\cal C}\left(\sigma_{1}^{+},\sigma_{1}^{-}\right)=1/(2\gamma)$,
${\cal C}\left(\sigma_{2}^{+},\sigma_{2}^{-}\right)=76/(63\gamma)$,
${\cal C}\left(\sigma_{2}^{+},\sigma_{1}^{-}\right)=22/(21\gamma)$,
${\cal C}\left(\sigma_{1}^{z},\sigma_{1}^{z}\right)=1/(32\gamma)$, ${\cal C}\left(\sigma_{2}^{z},\sigma_{2}^{z}\right)=1/(54\gamma)$,
${\cal C}\left(\sigma_{1}^{z},\sigma_{2}^{z}\right)=-1/(72\gamma)$
and ${\cal C}\left(\sigma_{2}^{z},\sigma_{1}^{z}\right)=1/(72\gamma)$.

The dominating term $\Gamma_{+}^{\text{ff}}{\cal D}[\cos\frac{\theta_{\text{ff}}}{2}S_{1}^{+}+\sin\frac{\theta_{\text{ff}}}{2}S_{2}^{+}]\rho$
[see the structure in Fig.~\ref{fig:rates} b)] possesses two stationary
states: $\left|\Psi_{\text{ss},1}\right>=\cos\frac{\theta_{\text{ff}}}{2} \left| \uparrow\downarrow \right>-\sin\frac{\theta_{\text{ff}}}{2} \left|\downarrow\uparrow\right>$ and $\left|\Psi_{\text{ss},2}\right>= \left| \uparrow\uparrow\right>$.
To make it unique, we can (i) add an extra channel or (ii) apply a coherent driving to the localized spins. 

\subsubsection{Two channels and no driving}
We introduce an extra channel
at the top with electrons flying in the opposite direction (from 4 to 3 in Fig.~\ref{fig:setup1}), opposite spin polarization and with the following symmetry in the exchange couplings: $J_{1}\equiv J_{1,1}=J_{2,4}$, $J_{2}\equiv J_{2,2}=J_{1,3}$.  
Summing up the first-order contributions from the two channels, the Zeeman energies (\ref{eq:ZeemanH})
necessary to cancel the first-order term are (see Eq.~\ref{eq:first-order-cancelation})
$\delta_{1(2)}=\mp\left(J_{1}\langle\sigma_{1}^{z}\rangle_{ss}-J_{2}\langle\sigma_{2}^{z}\rangle_{ss}\right)$ (the index in parentheses refers to the lower sign),
which in the case of equal rates become $\delta_{1(2)}=\pm\frac{J_{0}+7\delta J}{12}$.
% (see Fig.~\ref{fig:gradients}a)). 
%Zeeman energies $\delta_i$ of the order of
%the coupling strengths $J_i$ (i.e., typically a few $\mu$eV) can be readily
%achieved (e.g., in GaAs by magnetic fields of a few 100mT).

For the second-order term of the adiabatic elimination we need to calculate the correlation functions ${\cal C}\left(\sigma_{i}^{+},\sigma_{j}^{-}\right)$ and ${\cal C}\left(\sigma_{i}^{z},\sigma_{j}^{z}\right)$; $i,j=1,...4$; in particular this includes cross-correlations between the two channels. As the ancilla dot 4 (3) is symmetric to  1 (2), the correlations into the same channel do not need to be computed again. Since the ancilla quasisteady state does not contain any cross-channel correlations, nonlocal, cross-channel correlators vanish (when one traces out the ancilla degrees of freedom).
Then the new channel contributes mainly with the dissipator
$\Gamma_{+}^{\text{ff}}{\cal D}\left[\cos\frac{\theta_{\text{ff}}}{2}S_{2}^{-}+\sin\frac{\theta_{\text{ff}}}{2}S_{1}^{-}\right]\rho$
(note the symmetry $S_{1}^{+(-)}\leftrightarrow S_{2}^{-(+)}$)
and the effective ME for the system spins is
\begin{eqnarray}
\dot{\rho} & = & +\Gamma_{+}^{\text{ff}}{\cal D}\left[{\bf v}_{\text{ff}}^{+}\cdot\left(S_{1}^{+},S_{2}^{+}\right)\right]\rho\label{eq:meqeff1}\\
 & + & \Gamma_{+}^{\text{ff}}{\cal D}\left[{\bf v}_{\text{ff}}^{+}\cdot\left(S_{2}^{-},S_{1}^{-}\right)\right]\rho+{\cal L}_{\text{n-id}}^{(1)}\rho \ ,\nonumber 
\end{eqnarray}
where we have included all the non-dominating (nonideal) terms in
\begin{eqnarray}
{\cal L}_{\text{n-id}}^{(1)}\rho & = & -2\Delta\left[S_{2}^{-}S_{1}^{+}-S_{1}^{-}S_{2}^{+},\rho\right]\label{eq:Lnoise1} \\
 & + & \sum_{\text{\ensuremath{\sigma}=\ensuremath{\pm}}}\Gamma_{\sigma}^{\text{zz}}{\cal D}[{\bf v}_{\text{zz}}^{\text{\ensuremath{\sigma}}}\cdot\left(S_{1}^{z},S_{2}^{z}\right)]\rho\nonumber \\
 & + & \sum_{\text{\ensuremath{\sigma}=\ensuremath{\pm}}}\Gamma_{\sigma}^{\text{zz}}{\cal D}[{\bf v}_{\text{zz}}^{\text{\ensuremath{\sigma}}}\cdot\left(S_{2}^{z},S_{1}^{z}\right)]\rho\nonumber \\
 & + & \Gamma_{-}^{\text{ff}}{\cal D}[{\bf v}_{\text{ff}}^{\text{-}}\cdot\left(S_{1}^{+},S_{2}^{+}\right)]\rho+ \Gamma_{-}^{\text{ff}}{\cal D}[{\bf v}_{\text{ff}}^{\text{-}}\cdot\left(S_{2}^{-},S_{1}^{-}\right)]\rho\ .\nonumber
\end{eqnarray}
Here ${\bf v}_{\text{a}}^{+}=\left(\cos\frac{\theta_{\text{a}}}{2},\sin\frac{\theta_{\text{a}}}{2}\right)$
and ${\bf v}_{\text{a}}^{-}=\left(\sin\frac{\theta_{\text{a}}}{2},-\cos\frac{\theta_{\text{a}}}{2}\right)$, 
for $\text{a}=\text{ff},\text{zz}$.

\subsubsection{One channel and  driving}
 The second solution avoids the inclusion of a second channel and the
extra ancilla QDs and consists of applying a weak coherent driving field
in resonance with the Zeeman frequency, giving rise to the equation
\begin{eqnarray}
\dot{\rho} & = & -i\left[H_{\text{d}},\rho\right]-\Delta\left[S_{2}^{-}S_{1}^{+}-S_{1}^{-}S_{2}^{+},\rho\right]\nonumber \\
 & + & \Gamma_{+}^{\text{ff}}{\cal D}[{\bf v}_{\text{ff}}^{+}\cdot(S_{1}^{+},S_{2}^{+})]\rho+{\cal L}_{\text{n-id}}^{(2)}\rho\ ,\label{eq:meqeff2}
\end{eqnarray}
with the nonideal part
\begin{eqnarray}
{\cal L}_{\text{n-id}}^{(2)}\rho & = & \sum_{\text{\ensuremath{\sigma}=\ensuremath{\pm}}}\Gamma_{\sigma}^{\text{zz}}{\cal D}[{\bf v}_{\text{zz}}^{\text{\ensuremath{\sigma}}}\cdot\left(S_{1}^{z},S_{2}^{z}\right)]\rho\nonumber \\
 & + & \Gamma_{-}^{\text{ff}}{\cal D}[{\bf v}_{\text{ff}}^{\text{-}}\cdot\left(S_{1}^{+},S_{2}^{+}\right)]\rho\ .\label{eq:Lnoise2}
\end{eqnarray}
In this case, the Zeeman energies are $\delta_{i}=-J_{i}\langle\sigma_{i}^{z}\rangle_{ss}$.
%;
%see Fig.~\ref{fig:gradients}b).

\begin{figure}
\includegraphics[width=1\columnwidth]{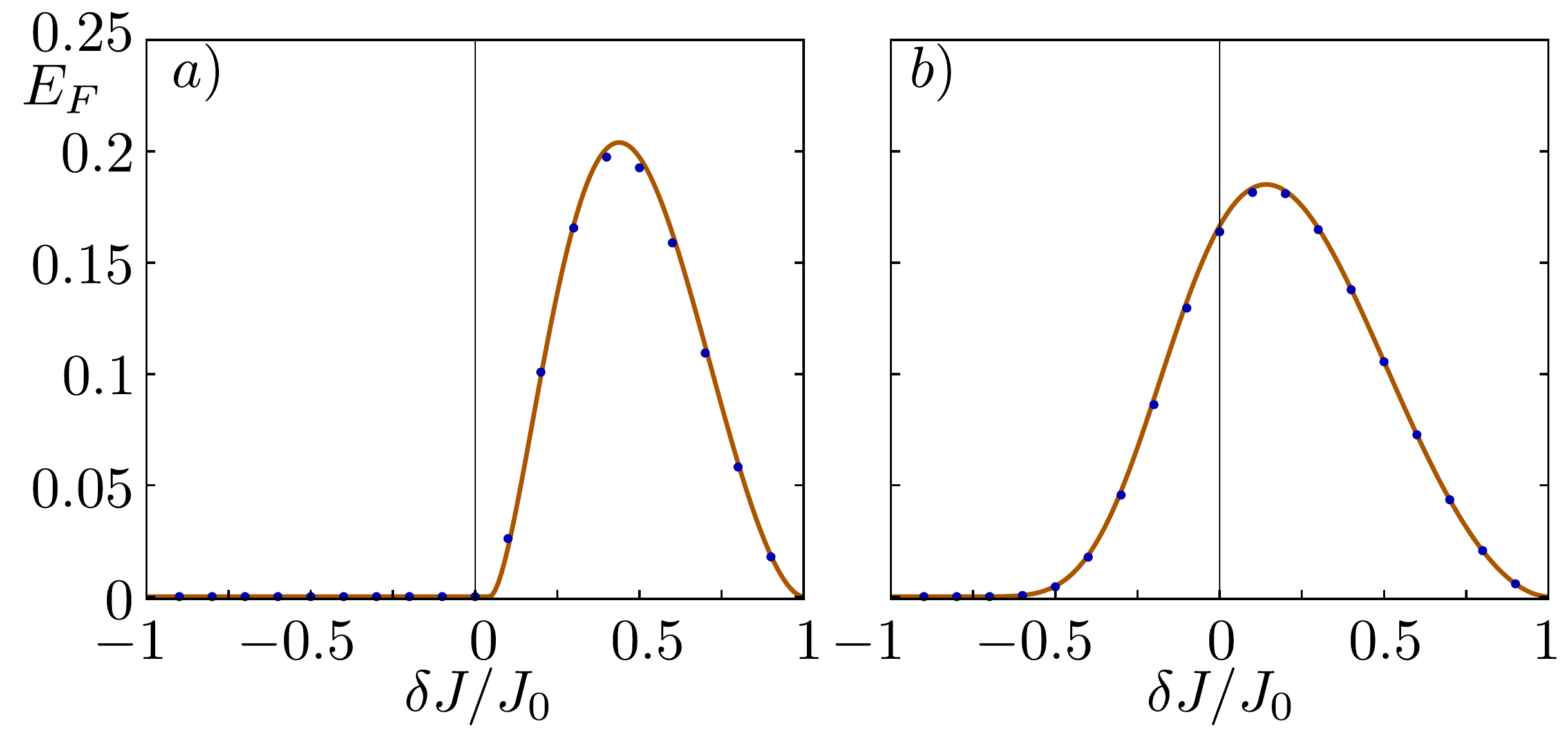}

\protect\caption{\label{fig:full-numerics}(color online). Steady-state entanglement  between two remote qubits
quantified via the $E_{F}$ for the two QHE-based proposals as a function of
$\delta J$.  
The solid lines in (a) and (b) refer to  Eq.~(\ref{eq:meqeff1}) and Eq.~(\ref{eq:meqeff2}), respectively, while the blue dots are calculated with the full ME including ancilla QDs in order to check the validity of our perturbative treatment.
Numerical parameters: $\gamma_{\text{L}}=\gamma=30\mu\text{eV}$, $J_{0}=3\mu\text{eV}$
and $\delta_{i} {\in}\left(-2,2\right)\mu\text{eV}$.
In (b), for each value of $\delta J$, 
$\Omega_{i}$ has been optimized in the range $\Omega_{i}{\in}\left(0-50\right)\text{neV}$.
%Comparison of the results before (blue dots)
%and after (orange line) adiabatic elimination  of the ancilla QDs. Proposal with a) 
%two channels and b) one channel and coherent driving. The figure depicts the $E_F$ of the steady state as a function
%of $\delta J$,
% with $J_{0}=3\mu\text{eV}$ and $\gamma_{\text{\text{L}}}=\gamma=30\mu\text{eV}$.
}
\end{figure}

%-----------------------------------------------------------------------------
\subsection{Validity of adiabatic elimination\label{sec:validity}}
In the main text, we discuss to what extent the entanglement of the localized spins inherent to the ideal dynamics persists despite the undesired  terms absorbed into ${\cal L}_{\text{n-id}}^{(i)}$. These results are based on the previous adiabatic elimination of ancilla dots.  
To check the validity of our perturbative treatment, in Fig.~\ref{fig:full-numerics} we compare the entanglement in
the steady state resulting from the full ME including ancilla
QDs to the Eqs.~(\ref{eq:meqeff1}) and ~(\ref{eq:meqeff2}),
i.e., after adiabatic elimination. For the experimentally achievable
parameters $\gamma=30\mu\text{eV}$ and $J_{0}=3\mu\text{eV}$
the  agreement is very good, 
%what confirms 
showing that the approximation is valid
for physically achievable conditions and it is possible to work with
the simplified effective ME for the system spins. Obviously,
the approximation becomes less accurate for larger values of the coupling
$J_{0}$ with respect to $\gamma$ (not shown).

\section{Effective Stroboscopic Evolution\label{sec:app-Effective-Stroboscopic-Evolution}}

In this appendix, we provide further details for the SAW-based setup explained in the main text. The protocol consists of a continuous train of mobile dots that interact successively with the two system spins. The concatenated evolution of the localized spins
DM is described by
\begin{eqnarray}
\rho^{(n)} & = & \text{tr}_{\text{a}}\left[ e^{{\cal L}_{2,n}\tau_{2}}e^{{\cal L}_{1,n}\tau_{1}}\left(\varrho^{(n-1)}\right)\right] \ ,\label{eq:rhon-1}\\
\varrho^{(n-1)} & = & \rho^{(n-1)}\otimes|\sigma_{n-1}\rangle\langle\sigma_{n-1}|
\end{eqnarray}
where $\rho^{(n)}$ defines the state of the system after the $n$-th
cycle of the protocol and the Liouvillian ${\cal L}_{i,n}$ encodes
the interaction of the ancilla electron with the system spin $i$
and the Zeeman Hamiltonian (\ref{eq:Zeeman}). Still, dephasing during transport could be included
straightforwardly in this model by adding a corresponding
super-operator in between the two interaction terms. For $J_i\tau_j,\delta_i\tau_j\ll1$, we can perform a short-time Taylor expansion  $e^{{\cal L}_{i,n}\tau}=1+\tau{\cal L}_{i,n}+\frac{\tau^{2}}{2}{\cal L}_{i,n}^{2}+...$ to approximate
$\rho^{(n)}$ to second order (let us employ for simplicity equal times $\tau\equiv\tau_{1}=\tau_{2}$)\begin{widetext}
\begin{eqnarray}
\rho^{(n)} & = & \text{tr}_{\text{a}}\left\{ \varrho^{(n-1)}-i\tau\left[H_{\text{Z}}+H_{\text{IN}}^{1,1},\varrho^{(n-1)}\right]-i\tau\left[H_{\text{Z}}+H_{\text{IN}}^{2,2},\varrho^{(n-1)}\right]\right\} \nonumber \\
 & + & \text{tr}_{\text{a}}\left\{ \frac{\tau^{2}}{2}{\cal D}\left[H_{\text{Z}}+H_{\text{IN}}^{1,1}\right]\varrho^{(n-1)}+\frac{\tau^{2}}{2}{\cal D}\left[H_{\text{Z}}+H_{\text{IN}}^{2,2}\right]\varrho^{(n-1)}\right\} \label{eq:general-step-2-1}\\
 & + & \tau^{2}\text{tr}_{\text{a}}\left\{ \left[H_{\text{Z}}+H_{\text{IN}}^{2,2},\varrho^{(n-1)}\left(H_{\text{Z}}+H_{\text{IN}}^{1,1}\right)\right]+\left[\left(H_{\text{Z}}+H_{\text{IN}}^{1,1}\right)\varrho^{(n-1)},H_{\text{Z}}+H_{\text{IN}}^{2,2}\right]\right\} +{\cal O}\left(\tau^{3}J^{3}\right)\ .\nonumber 
\end{eqnarray}
When the injected spin is $\left|\sigma_{n-1}\right\rangle =\left|\uparrow\right\rangle $,
\begin{eqnarray}
\rho^{(n)} & = & \rho^{(n-1)}-2i\tau\left[\delta_{1}S_{1}^{z}+\delta_{2}S_{2}^{z},\rho^{(n-1)}\right]-\frac{i}{2}\tau\left[J_{1}^{\uparrow}S_{1}^{z}+J_{2}^{\uparrow}S_{2}^{z},\rho^{(n-1)}\right]\label{eq:up}\\
 & + & \frac{\tau^{2}}{2}{\cal D}\left[\left(2\delta_{1}+\frac{J_{1}^{\uparrow}}{2}\right)S_{1}^{z}+\left(2\delta_{2}+\frac{J_{2}^{\uparrow}}{2}\right)S_{2}^{z}\right]\rho^{(n-1)}\nonumber \\
 & + & \frac{1}{8}{\cal D}\left[\tau J_{1}^{\uparrow}S_{1}^{+}+\tau J_{2}^{\uparrow}S_{2}^{+}\right]\rho^{(n-1)}+\tau^{2}\frac{J_{1}^{\uparrow}J_{2}^{\uparrow}}{8}\left[S_{1}^{-}S_{2}^{+}-S_{2}^{-}S_{1}^{+},\rho^{(n-1)}\right]+{\cal O}\left(\tau^{3}J^{3}\right)\ \nonumber 
\end{eqnarray}
and if $\left|\sigma_{n}\right\rangle =\left|\downarrow\right\rangle$ the next step is given by
\begin{eqnarray}
\rho^{(n+1)} & \simeq & \rho^{(n)}-2i\tau\left[\delta_{1}S_{1}^{z}+\delta_{2}S_{2}^{z},\rho^{(n)}\right]+\frac{i}{2}\tau\left[J_{1}^{\downarrow}S_{1}^{z}+J_{2}^{\downarrow}S_{2}^{z},\rho^{(n)}\right]\label{eq:down}\\
 & + & \frac{\tau^{2}}{2}{\cal D}\left[\left(2\delta_{1}-\frac{J_{1}^{\downarrow}}{2}\right)S_{1}^{z}+\left(2\delta_{2}-\frac{J_{2}^{\downarrow}}{2}\right)S_{2}^{z}\right]\rho^{(n)}\nonumber \\
 & + & \frac{1}{8}{\cal D}\left[\tau J_{1}^{\downarrow}S_{1}^{-}+\tau J_{2}^{\downarrow}S_{2}^{-}\right]\rho^{(n)}-\tau^{2}\frac{J_{1}^{\downarrow}J_{2}^{\downarrow}}{8}\left[S_{1}^{-}S_{2}^{+}-S_{2}^{-}S_{1}^{+},\rho^{(n)}\right]+{\cal O}\left(\tau^{3}J^{3}\right)\ .\nonumber
\end{eqnarray}
\end{widetext}Analogously to the two proposals of the QHE-based setup, we consider (i) a protocol with alternating spin directions and suitably synchronized exchange couplings and (ii) a spin-polarized protocol with a coherent driving. Both transport protocols drive the localized spins to an entangled state independent of the initial state.

\subsubsection{Alternating spin sequences}
The concatenation of two steps with the injection of an opposite
spin results in a first-order term that can be canceled by choosing the
Zeeman energies as $\delta_{i}=-\frac{ J_{i}^{\uparrow}- J_{i}^{\downarrow}}{8}$. Setting in addition $\tau
J_{1}^{\uparrow} = \tau J_{2}^{\downarrow}\equiv \mu$ and $\tau J_{1}^{\downarrow} =
\tau J_{2}^{\uparrow}\equiv\nu$, this is simply a gradient of magnetic field
between the two localized spins: $\delta_{1(2)}=\mp\frac{\delta J}{4}$, with $J_{1(2)}^{\uparrow}=J_0\mp\delta J$.
%; see Fig.~\ref{fig:gradients-SAWs} a). 
Not only the first-order terms but also the dephasing second-order terms
in Eqs.~(\ref{eq:up}) and~(\ref{eq:down}) cancel and it is readily seen that
\begin{eqnarray}
\rho^{(n+1)} & = & \rho^{(n-1)}+\frac{1}{8}{\cal D}\left[\mu S_{1}^{+}+\nu S_{2}^{+}\right]\rho^{(n-1)}\label{eq:n+1-n-1-simple}\\
 & + & \frac{1}{8}{\cal D}\left[\nu S_{1}^{-}+\mu S_{2}^{-}\right]\rho^{(n-1)}+{\cal O}\left(\tau^{3}J^{3}\right)\ ,\nonumber 
 \end{eqnarray}
whose second-order terms are the considered ideal dynamics in the main text because they have a unique pure entangled steady state.

%The ideal, analytical result given in Eq.~(\ref{eq:n+1-n-1-simple}) assumes the injection of alternating spin components of the form $\uparrow, \downarrow, \uparrow, \dots$. However, this condition can be relaxed to longer sequences of aligned ancilla spins, of the form $\uparrow, \uparrow, \dots, \downarrow, \downarrow,\dots,\uparrow,\uparrow,\dots$. This has been confirmed numerically in Fig.~\ref{fig:app-strob-Passive-scheme-with-mobiledots}. Accordingly, the switching times of the gates can be increased by about an order of magnitude without severely affecting the amount of steady-state entanglement.     

\subsubsection{Single spin-component and driving}
 For the protocol with a single spin component the approximated stroboscopic evolution
is given by Eq.~(\ref{eq:up}), therefore by choosing the magnetic
fields with strengths $\delta_{i}=-J_{i}^{\uparrow}/4$, 
we cancel the first-order contribution.
%; see Fig.~\ref{fig:gradients-SAWs} b). 
With the definitions $\mu=J_{1}^{\uparrow}\tau$
and $\nu=J_{2}^{\uparrow}\tau$ and applying a coherent driving
\begin{equation}
H_{\text{d}}=\sum_{i=1,2}2\Omega_{i}S_{i}^{x}\label{eq:Hdriving}
\end{equation}
such that $\Omega_{1,2}\ll J$, the stroboscopic evolution reads
\begin{eqnarray}
\rho^{(n)} & = & \rho^{(n-1)}-2i\tau\left[H_{\text{d}},\rho^{(n-1)}\right]\label{eq:n-n-1-driving}\\
 & + & \frac{\mu\nu}{8}\left[S_{1}^{-}S_{2}^{+}-S_{2}^{-}S_{1}^{+},\rho^{(n-1)}\right]\nonumber \\
 & + & \frac{1}{8}{\cal D}\left[\mu S_{1}^{+}+\nu S_{2}^{+}\right]\rho^{(n-1)}+{\cal O}\left(\tau^{3}J^{3}\right)\ ,\nonumber 
\end{eqnarray}
which is also like the desired one up to second order.

Note that for a direct comparison of Eqs.~(\ref{eq:n+1-n-1-simple}) and~(\ref{eq:n-n-1-driving}) with a ME, one needs to assume infinitesimal interaction times, but we have confirmed that the schemes work for finite interaction times.

%-----------------------------------------------------------------------------
\section{Noise Sources\label{sec:ap-Noise-sources}}

In this appendix we detail the different noise sources taken into account in the proposed setups. 
First of all, we account for qubit dephasing induced by nuclear spins in the (GaAs) host environment. 
Second, we consider electron losses due to imperfections in the transport mechanisms. 
Then, we analyze the effect of an imperfect cancellation of the first-order terms, i.e., the effect of some residual gradient.
Finally, in the SAW-based proposal we account for imperfections due to uncertainties in the effective electron interaction times.

To account for dephasing due to the nuclear spins, we follow the
standard treatment \citep{Chekhovich2013} and assume that the spins
in the QDs experience non-Markovian noise. The fluctuations
of the  Overhauser field lead to a time-ensemble-averaged
electron dephasing time $T_{2}^{*}$, that is related to the width
of the nuclear field distribution $\sigma_{\text{nuc}}$ as $T_{2}^{*}=\sqrt{2}/\sigma_{\text{nuc}}$.
In order to model this effect, we have to include the Hamiltonian
\citep{Taylor2007,Kloeffel2013,Chekhovich2013}
\begin{equation}
H_{\text{deph}}=\sum_{i=1,2}B_{i}^{\text{nuc}}S_{i}^{\text{z}}\label{eq:Ldeph}
\end{equation}
with random parameters $B_{i}^{\text{nuc}}$ sampled independently from a
normal distribution with standard deviation $\sigma_{\text{nuc}}$. 

%spin-echo
Before we proceed, we note that, due to the several time scales involved, our scheme should be very amenable to the inclusion of dynamical decoupling techniques, which allow for significantly extended electron coherence times, $T_2 \approx 10^{2}T_{2}^{*}$ \citep{Chekhovich2013, Kloeffel2013, Taylor2007}.

%-----------------------------------------------------------------------------
\subsection{Transport via QHE states}
\label{sec:noiseQHE}
 
The full MEs
were derived in Appendix~\ref{sec:Effective-master-equation}. In Fig.~\ref{fig:coop} we plot the $E_F$ of the steady state for different
values of the 9-like parameter $C=J_{0}^{2}/\gamma\sigma_{\text{nuc}}$, which compares desired $\sim J_0^2/\gamma$ to undesired $\sim \sigma_{\text{nuc}}\sim 1/T_{2}^{*}$ rates. 
As expected from the analysis of the
spectral gap 
%in Section~\ref{sec:Effective-master-equation}, 
the
purely dissipative proposal is typically found to be more robust.
By choosing the values
$\gamma=30\mu\text{eV}$ and $J_{0}=3\mu\text{eV}$ we can predict
that a value of $\sigma_{\text{nuc}}=0.03\mu\text{eV}$, which corresponds
to a cooperativity $C=10$, would be very good concerning the purely
dissipative proposal. This standard deviation corresponds to a dephasing
time $T_{2}^{*}\simeq30\text{ns}$, which is experimentally feasible
and can be improved up to $3\mu\text{s}$ using nuclear-state-narrowing
techniques \citep{Shulman2014,Chekhovich2013}. 

\begin{figure}
\includegraphics[width=1\columnwidth]{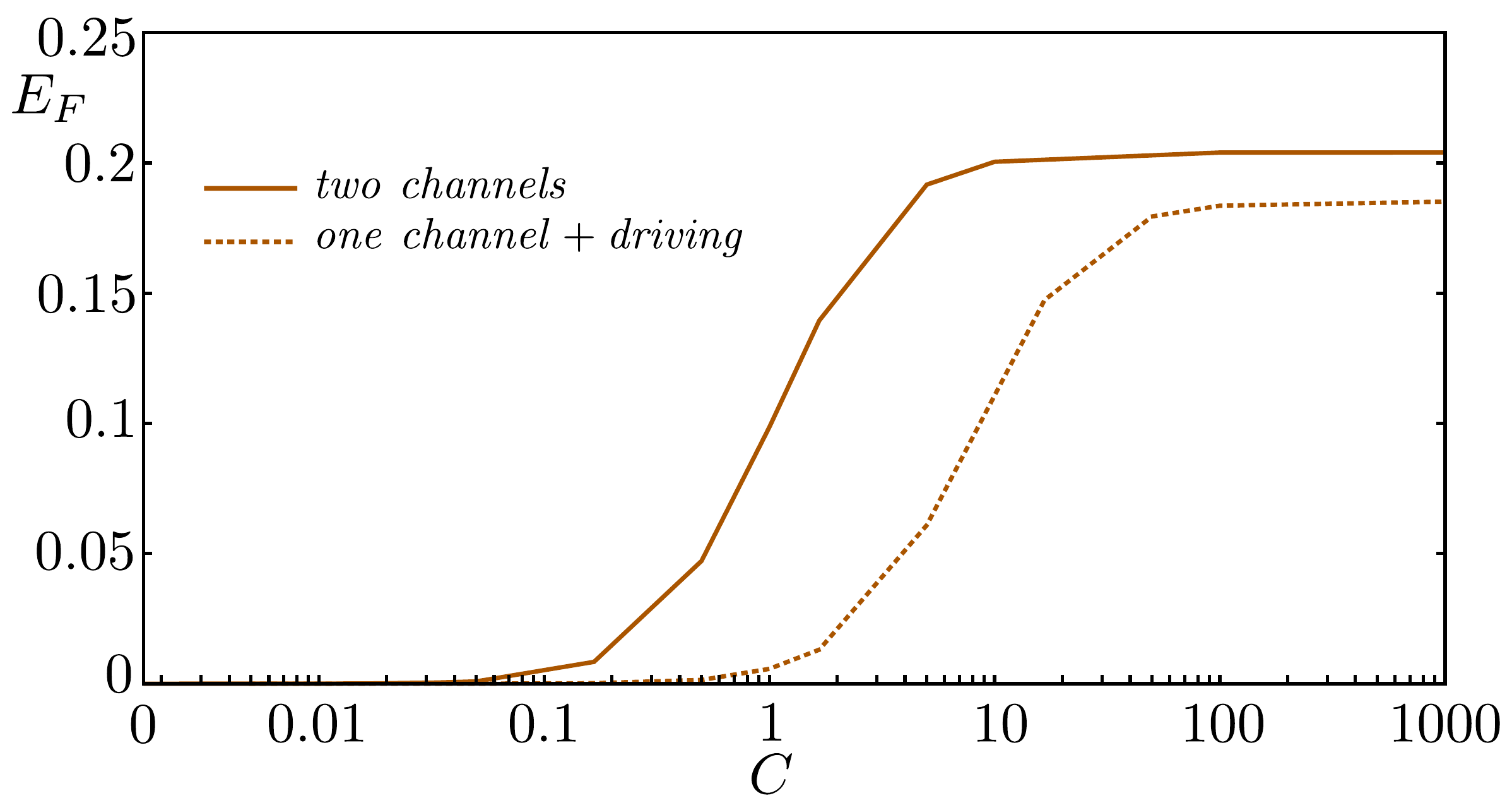}

\protect\caption{\label{fig:coop}Steady-state entanglement   between two remote qubits
quantified via the $E_{F}$ for the two QHE-based proposals as a function
of the cooperativity $C=J_{0}^{2}/\gamma\sigma_{\text{nuc}}$.
The solid (dotted) line 
results are based on Eq.~(\ref{eq:meqeff1}) and Eq.~(\ref{eq:meqeff2}), respectively. 
Numerical parameters: $\gamma_{\text{L}}=\gamma=30\mu\text{eV}$, $J_{0}=3\mu\text{eV}$, 
$\delta J/J_{0}=0.44$ ($\delta J/J_{0}=0.14$) for solid (dotted) line 
and $\delta_{i} {\in}\left(-2,2\right)\mu\text{eV}$.
% Steady-state entanglement between two remote qubits quantified via the $E_F$ for the two QHE-based proposals as a function
% Continuous
%line) purely dissipative proposal with two channels ($\delta J/J_{0}=0.44$).
%Dotted line) proposal with one channel and coherent driving ($\delta J/J_{0}=0.14$). 
}
\end{figure}

\begin{figure}[h]
\includegraphics[width=1\columnwidth]{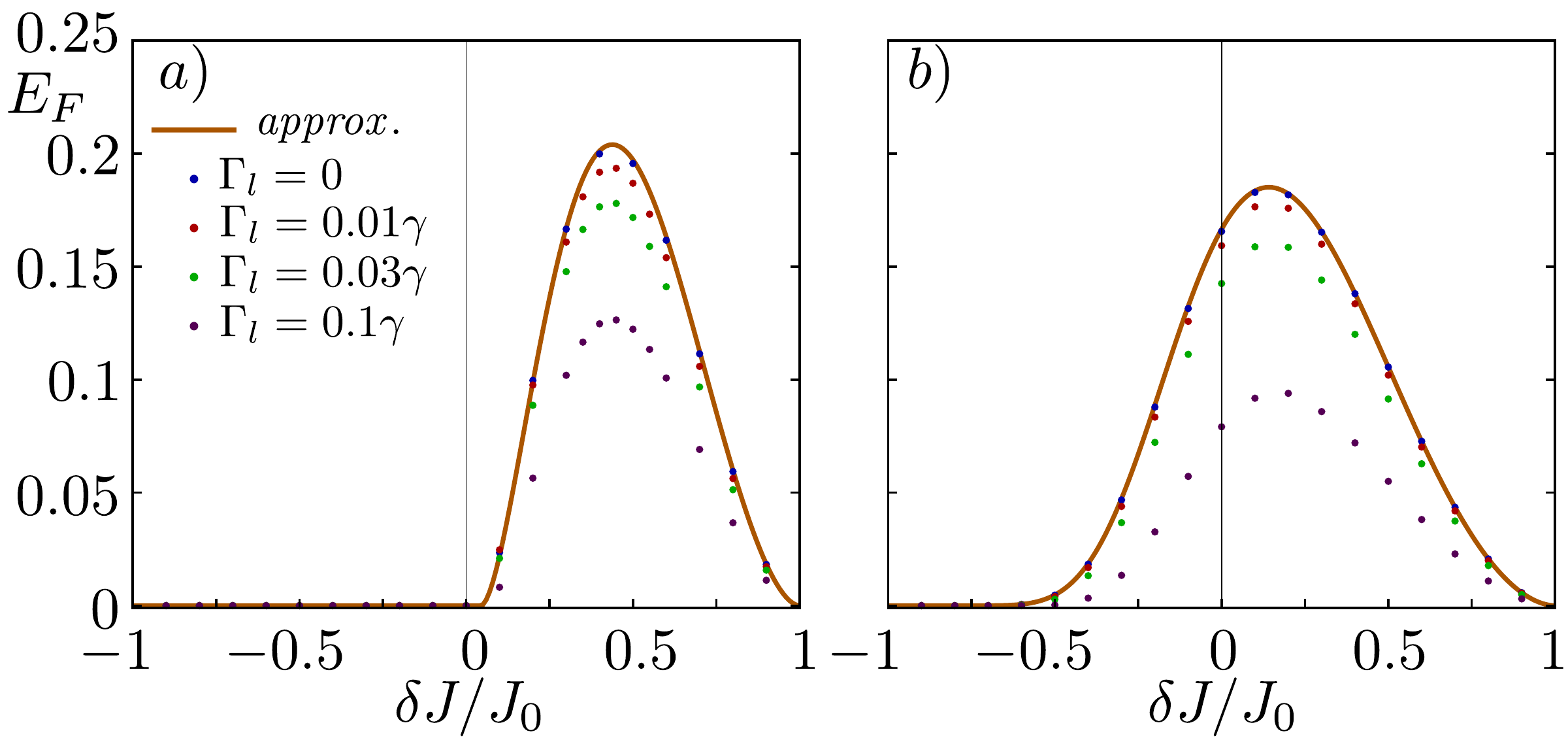}

\protect\caption{\label{fig:losses-QHE}(color online). Steady-state entanglement  between two remote qubits
quantified via the $E_{F}$ for the two QHE-based proposals as a function of
$\delta J$. The solid lines in (a) and (b) refer to  Eq.~(\ref{eq:meqeff1}) and Eq.~(\ref{eq:meqeff2}), respectively, while the dots are calculated with the full ME including ancilla QDs and different losses rates 
%a) (b))
%results are based on Eq.~(\ref{eq:meqeff1}) and Eq.~(\ref{eq:meqeff2}), respectively. 
%The continuous line corresponds to the approximated result with adiabatic elimination of the ancilla system and the dots correspond to different losses rates
 $\Gamma_{l}$.
Numerical parameters: $\gamma_{\text{L}}=\gamma=30\mu\text{eV}$, $J_{0}=3\mu\text{eV}$
and $\delta_{i} {\in}\left(-2,2\right)\mu\text{eV}$.
In (b), for each value of $\delta J$, 
$\Omega_{i}$ has been optimized in the range $\Omega_{i}{\in}\left(0-50\right)\text{neV}$.
%Steady-state entanglement between two remote qubits quantified via the $E_F$ for the two QHE-based proposal with a) 
%two channels and b) one channel and coherent driving as a function
%of $\delta J$, with $J_{0}=3\mu\text{eV}$ and $\gamma_{\text{\text{L}}}=\gamma=30\mu\text{eV}$.
}
\end{figure}

To model the possible electron losses due to imperfections in the transport channel, we include a Lindblad operator  with rate $\Gamma_l$ acting in the first ancilla QD, i.e., $\sum_{\sigma}\Gamma_l/2{\cal{D}}\left[d_1^{\sigma}\right]$ (also in $d_4^{\sigma}$ in the two-channels proposal). The result, shown in Fig.~\ref{fig:losses-QHE}, predicts that we can afford a small percent of losses.

\begin{figure}[h]
\includegraphics[width=1\columnwidth]{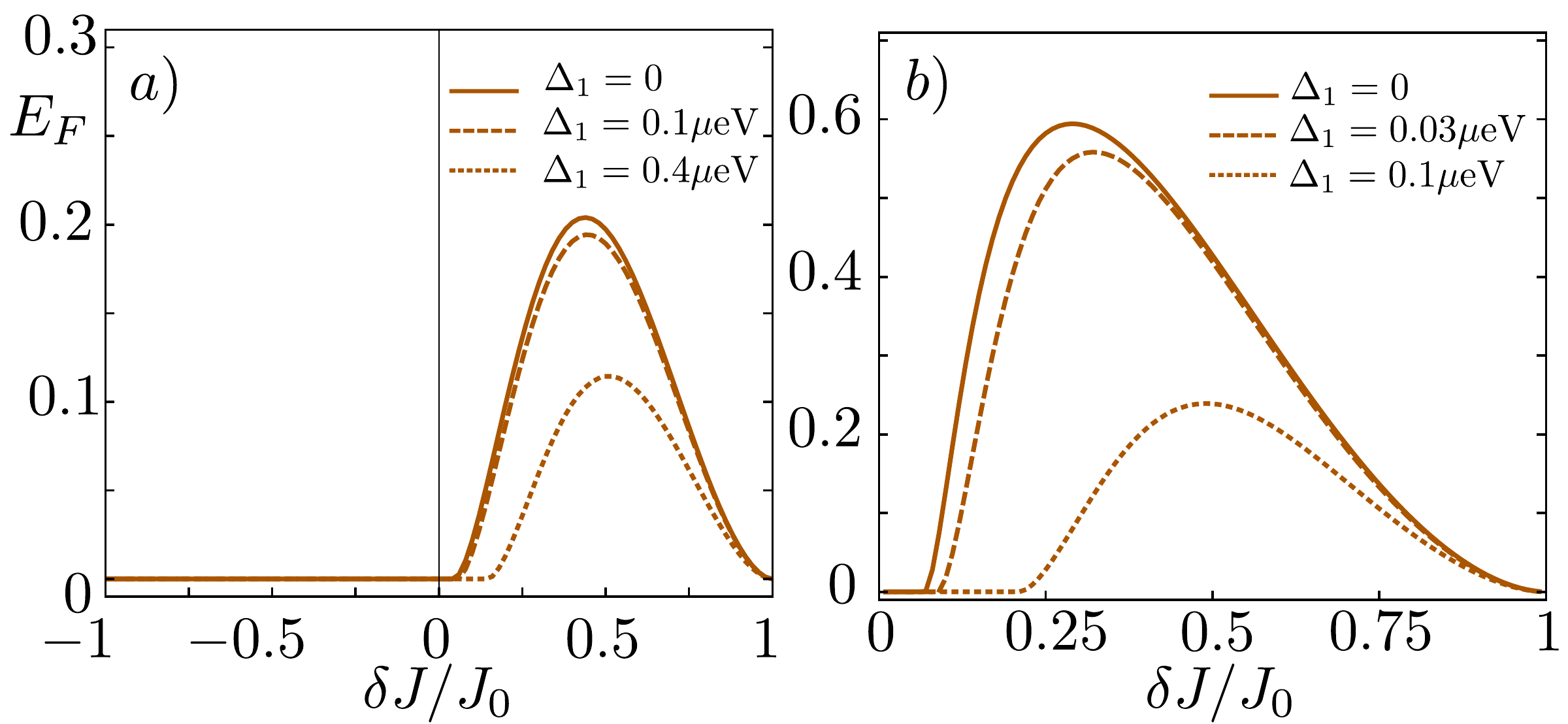}

\protect\caption{\label{fig:no-cancellation} Steady-state entanglement between two remote qubits quantified via the $E_F$ for 
two proposals as a function of
$\delta J$. 
The solid lines in (a) and (b) refer to  Eq.~(\ref{eq:meqeff1}) and Eq.~(\ref{eq:n+1-n-1-simple}), respectively, while the results in dashed and dotted lines account for 
%a) QHE-based proposal with  
%two channels and b) SAWs-based proposal with alternating sequences as a function
%of $\delta J$ for 
different values of the residual gradient $\Delta_1$ ($\Delta_2=0$). Numerical parameters: (a) $\gamma_{\text{L}}=\gamma=30\mu\text{eV}$, $J_{0}=3\mu\text{eV}$. (b)  $J_{0}=2.5\mu\text{eV}$, $\tau=0.1\text{ns}$.}
\end{figure}

Finally, we verify in Fig.~\ref{fig:no-cancellation} (a) that the perfect cancellation of the first-order terms is not necessary, provided that the residual gradients $\Delta_{i}$ are small compared to the gap.

%-----------------------------------------------------------------------------
\subsection{Transport via SAW moving dots}
\label{sec:noisesaws}
The approximated
Eqs.~(\ref{eq:n+1-n-1-simple}) and~(\ref{eq:n-n-1-driving})
suggest that the simulation of the full problem given in Eq.~(\ref{eq:rhon-1})
will drive the main qubits to an entangled steady state regardless of the initial state (as long as $\tau J_{i}\ll1$). However, in a realistic experimental
situation, there will be also some noise sources. In the following,
we account for: (i) dephasing due to the
nuclear spins, (ii) imperfections due to the uncertainty in the
dwell time $\tau$ (time jitter), (iii) electron losses due to imperfections in the transport mechanism and (iv) residual gradients. (i) As explained above, we include a dephasing Hamiltonian as in Eq.~(\ref{eq:Ldeph}) to model the non-Markovian
noise due to the hyperfine interaction. We
assume that the ancilla dots are refilled very quickly after every step
and thus neglect the evolution in the short intermittent intervals when
the ancilla dot is empty. (ii) In a realistic experimental
situation, there will be also some noise associated with the uncertainty
in the dwell times \citep{Bocquillon2013}. We include this noise
source by choosing the times $\tau_{i}$ randomly from a Gaussian
distribution centered around the average ($\tau$) with a
standard deviation of $\sigma_{\tau}$. (iii) To model the losses we assume during the time simulation that  with  a certain probability an ancilla spin never interacts with the second localized spin. (iv) We estimate how large the imperfections in the magnetic gradients can be such that the entanglement generation is not severely affected.

\begin{figure}
\includegraphics[width=1\columnwidth]{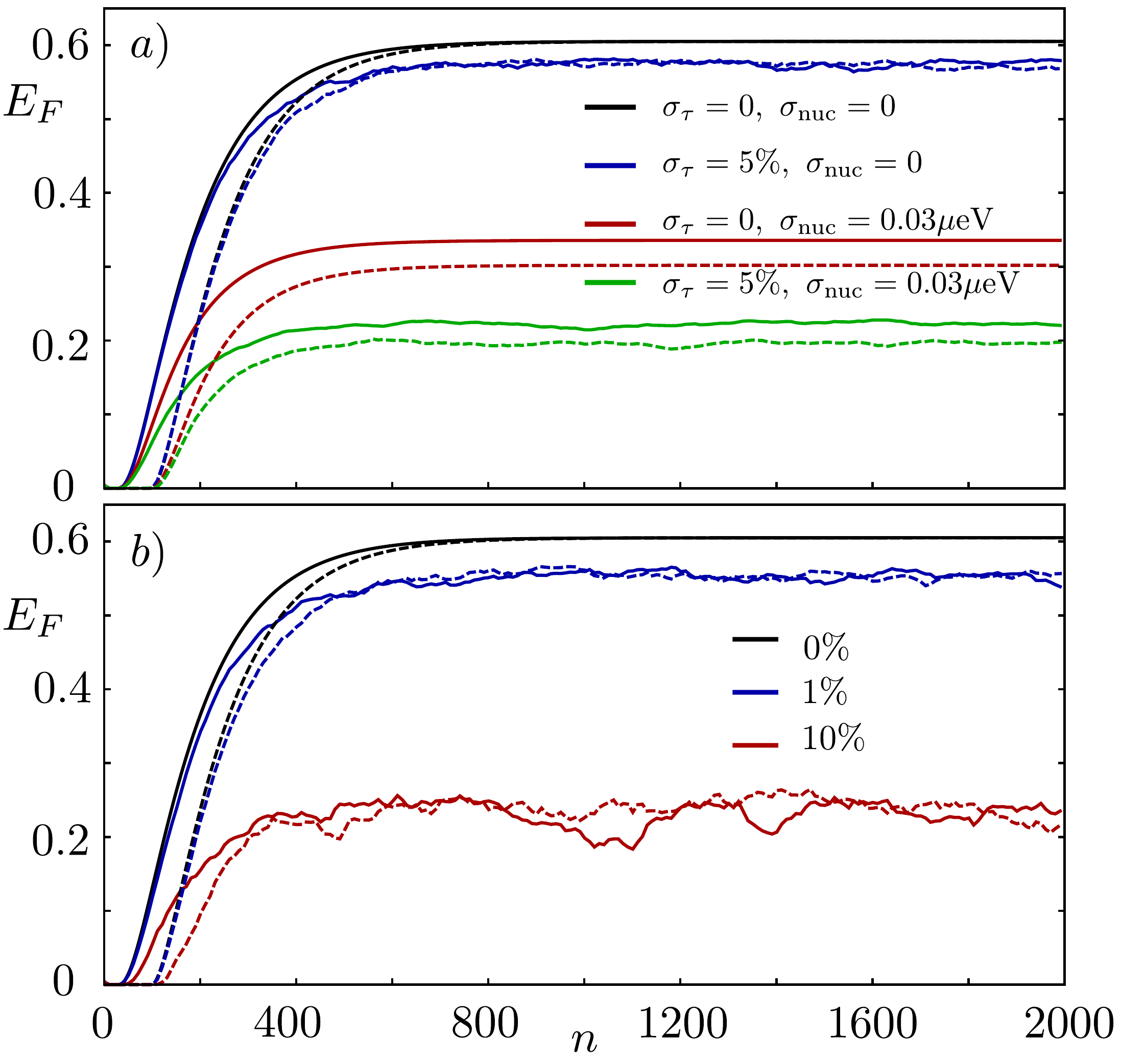}
\protect\caption{\label{fig:app-noise-Passive-scheme-with-mobiledots}
(color online). Entanglement  between two remote qubits quantified via
 the $E_F$ for the  SAW-based proposal corresponding to
 Eq.~(\ref{eq:n+1-n-1-simple}) as a function of time ($t=2\text{n}\tau$)
 for two different initial states (solid and dashed
 lines, respectively) and $\delta J/J_{0}=0.28$,
 $J_{0}=2.5\mu\text{eV}$ and $\tau=0.1\text{ns}$. In both (a) and (b), the
 black curves depict the ideal case and the remaining curves show the
 effect of different kinds of noise (time jitter $\sigma_{\tau}$ and
 nuclear dephasing in (a); electron losses in (b) averaged
 over several random trajectories of the respective processes.}
\end{figure}

In Fig.~\ref{fig:app-noise-Passive-scheme-with-mobiledots} we show the effect of the  noise sources (i), (ii) and (iii) in the simulation
in terms of $E_F$ of the state. The convergence is found after $\sim10^{3}$ iterations,
which corresponds to the regime of $(0.1-1)\mu\text{s}$ for $\tau=(0.1-1)\text{ns}$.
Note that if the product $J_{0}\tau$ is fixed, the results do not change, but the time to reach the steady state and consequently the undesired dephasing  decrease with $\tau$. Once a small enough $\tau$ is fixed, the result improves as $J_0$ decreases but obviously the time grows and we need to find a compromise between the conditions $\tau J_0\ll 1$ and a time sufficiently short for the given nuclear dephasing time.
In Fig.~\ref{fig:no-cancellation} b) we show the effect of (iv) in the entanglement generation scheme with alternating spins.

\begin{figure}[h]
\includegraphics[width=1\columnwidth]{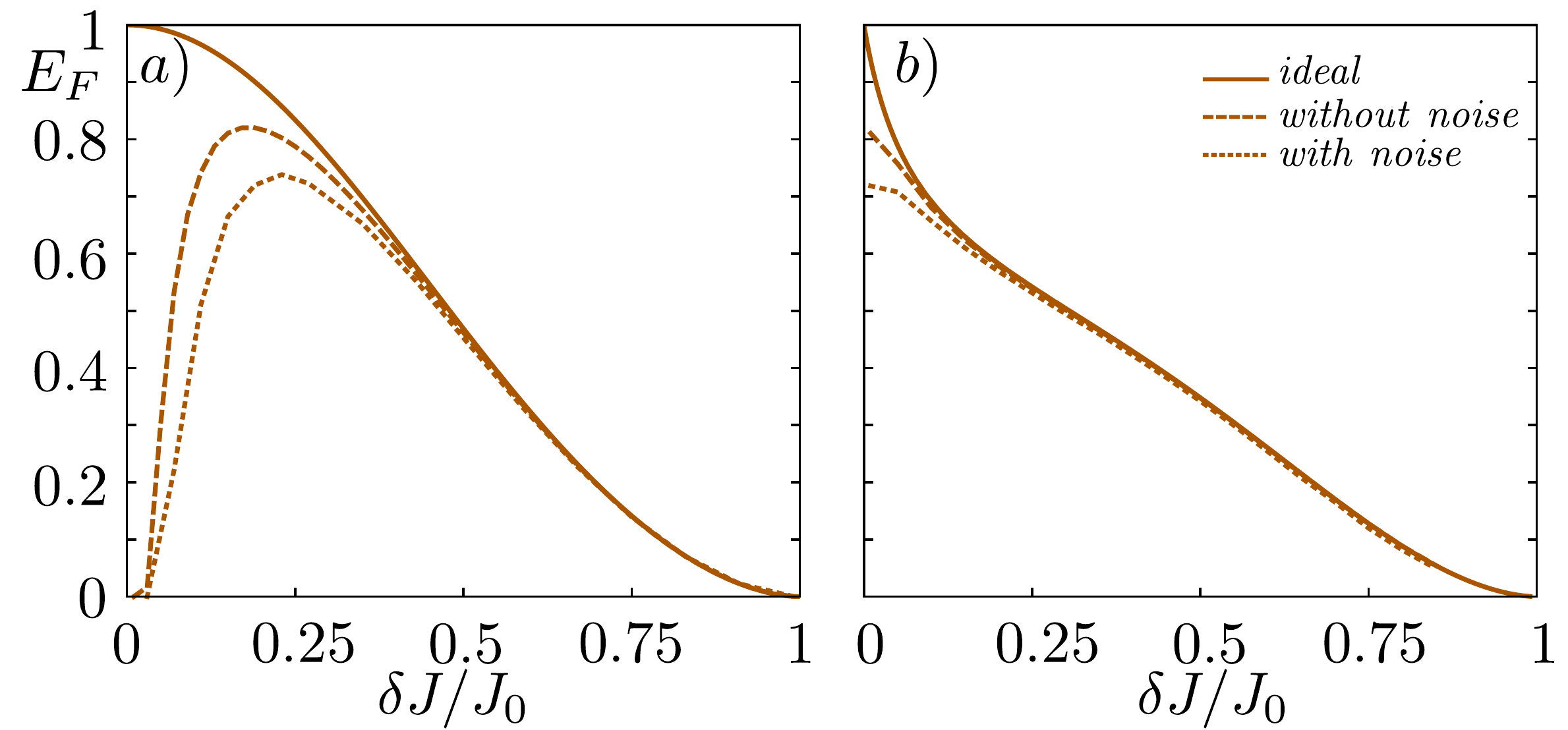}

\protect\caption{\label{fig:optimistic-case}Steady-state entanglement between two remote qubits quantified via the $E_F$ for the two SAW-based proposals as a function of
$\delta J$ ($J_{1(2)}^{\uparrow}=J_{0}\mp\delta J$). 
(a) and (b) show the results of Eq.~(\ref{eq:n+1-n-1-simple}) and Eq.~(\ref{eq:n-n-1-driving}), respectively.
%a) Alternating spin sequences with synchronized couplings, Eq.~(\ref{eq:n+1-n-1-simple}). 
%b) Spin-filtered ancilla spins and coherent driving, Eq.~(\ref{eq:n-n-1-driving}). 
The solid lines refer to the ideal result, given by the lower order terms present in Eqs.~(\ref{eq:n+1-n-1-simple}) and~(\ref{eq:n-n-1-driving}), while the dashed lines correspond to the full evolution. 
The dotted lines also account for noise due to uncertainty in the dwell times and dephasing.
Numerical parameters: $\sigma_{\text{\ensuremath{\tau}}}=5\%$, $J_{0}\tau{\approx}0.15$ and
$T_{2}^{*}/\tau{\approx}30000$. In (b), for each value of $\delta J$, 
$\Omega_{i}$ 
has been optimized in the range  $\Omega_{i}\tau\in\left(0-3\right)\cdot 10^{-3}$.
}
\end{figure}

The short dephasing times considered within the main text force us to choose a quite large value of $\tau J_0=0.38$; therefore the amount of entanglement generated is bounded to $E_F\gtrsim0.4$. 
 If the dephasing time reaches the maximal
experimental reported value of $T_{2}^{*}=3\mu\text{s}$, the amount
of steady-state entanglement increases up to $E_F\gtrsim0.7$, as shown in Fig.~\ref{fig:optimistic-case}.

%-----------------------------------------------------------------------------
%-----------------------------------------------------------------------------
%-----------------------------------------------------------------------------
%-----------------------------------------------------------------------------
%-----------------------------------------------------------------------------

%\bibliography{ref-entanglement-paper2,ref-entanglement-paper,ref-entanglement-paper3}

%

\end{document}